\RequirePackage{fix-cm}
\documentclass{svjour3}
\usepackage{graphicx}
\usepackage{hyperref}
\usepackage[round]{natbib}
\smartqed

\begin{document}
\title{Pore-network stitching method: A pore-to-core upscaling approach for multiphase flow}
\author{Amir H. Kohanpur \and Albert J. Valocchi}
\institute{A. H. Kohanpur \at
              University of Illinois at Urbana-Champaign \\
              \email{kohanpu2@illinois.edu}
           \and
           A. J. Valocchi \at
              University of Illinois at Urbana-Champaign \\
              \email{valocchi@illinois.edu}
}
\date{Received: date / Accepted: date}
\maketitle

\begin{abstract}
Pore-network modeling is a widely used predictive tool for pore-scale studies in various applications that deal with multiphase flow in heterogeneous natural rocks. Despite recent improvements to enable pore-network modeling on simplified pore geometry extracted from rock core images and and its computational efficiency compared to direct numerical simulation methods, there are still limitations to modeling a large representative pore-network for heterogeneous cores. These are due to the technical limits on sample size to discern void space during X-ray scanning and computational limits on pore-network extraction algorithms. Thus, there is a need for pore-scale modeling approaches that have the natural advantages of pore-network modeling and can overcome these limitations, thereby enabling better representation of heterogeneity of the 3D complex pore structure and enhancing the accuracy of prediction of macroscopic properties. This paper addresses these issues with a workflow that includes a novel pore-network stitching method to provide large-enough representative pore-network for a core. This workflow uses micro-CT images of heterogeneous reservoir rock cores at different resolutions to characterize the pore structure in order to select few signature parts of the core and extract their equivalent pore-network models. The space between these signature pore-networks is filled by using their statistics to generate realizations of pore-networks which are then connected together using a deterministic layered stitching method. The output of this workflow is a large pore-network that can be used in any flow and transport solver. We validate all steps of this method on different types of natural rocks based on single-phase and two-phase flow properties such as drainage relative permeability curves of carbon dioxide and brine flow. Then, we apply the stochastic workflow on two large domain problems, connecting distant pore-networks and modeling a heterogeneous core. We generate multiple realizations and compare the average results with properties from a defined reference pore-network for each problem. We demonstrate that signature parts of a heterogeneous core, which are a small portion of its entire volume, are sufficient inputs for the developed pore-network stitching method to construct a representative pore-network and predict flow properties.
\end{abstract}


\section{Introduction}
\label{intro}
The physics of two-phase flows in heterogeneous natural rocks is relevant to many applications of multiphase flows in porous media such as geological sequestration of carbon dioxide (CO\textsubscript{2}), displacements in oil recovery, and contaminant transport. Pore-scale modeling is essential in understanding fundamental phenomena necessary for prediction of macroscopic properties used in field-scale simulation tools. As an example, relative permeability and capillary pressure curves are important inputs of field-scale models to predict long-term displacement of oil or trapping of CO\textsubscript{2} in reservoirs. Although these properties can be estimated from core-scale experiments, these laboratory experiments are difficult to implement, time-consuming, expensive, and have uncertainties in measurements (\citealt{joekar_2012}). Prediction of macroscopic flow and transport properties of rocks based on pore-scale physics has been a focus of many  studies (\citealt{blunt_2013}). Pore-scale models are generally less expensive than experiments and have more flexibility in implementing and changing parameters.

With recent advances in instrumentation and X-ray computed tomography (CT) imaging, one can directly obtain detailed three-dimensional (3D) geometry of rock and its pore structure with high resolution images (\citealt{flannery_1987}; \citealt{dunsmuir_1991}; \citealt{wildenschild_2013}; \citealt{andra_2013_i}; \citealt{andra_2013_ii}). Such images are inputs of pore-scale models which can be categorized as either direct or simplified models, based on the manner in which they represent the geometry of the void space of porous media. In direct models, multiphase flow and transport equations are solved directly on the exact geometry of pore space obtained from the CT images. Some examples of direct numerical simulation (DNS) methods are lattice-Boltzmann methods (\citealt{boek_2010}; \citealt{ramstad_2012}), finite volume methods (\citealt{ferrari_2013}; \citealt{raeini_2014}), level set methods (\citealt{prodanovic_2006}), etc. The main challenges in using DNS methods are limitations on the size of the domain that is relevant to the grids resolution of simulation and limitations on computational efficiency.

On the other hand, a popular simplified model is pore-network (PN) modeling (\citealt{fatt_1956}; \citealt{sahimi_2011}; \citealt{blunt_2017}) which simplifies the pore space by dividing it into two categories of pore elements: pore-bodies (larger elements for storage of fluids) and pore-throats (narrower elements for flow of fluids). A PN can be either generated from statistics of the pore space, such as pore size distribution, or extracted directly from the 3D image of rock where irregular surfaces and edges of the pore space are abstracted down to simpler geometrical units. There are different PN extraction algorithms in the literature such as medial axis (\citealt{lindquist_1996}), watershed (\citealt{gostick_2017}), and maximal ball (MB) algorithms (\citealt{silin_2006};\citealt{dong_2009}). Defining a PN requires geometrical (location, size, and shape of pore elements) and topological (connections between pore elements) information of the pore space. PN flow models then use some assumptions and approximations to the governing equations on the entire PN e.g., Hagen-Poiseuille equation for flow in pore-throats. More details can be found in \cite{valvatne_2004} and \cite{joekar_2012}. These approximations to geometry and physics make PN models more capable of simulating a representative elementary volume (REV) with less computational effort compare to DNS methods to predict macroscopic properties.

Although PN modeling based on CT images has been applied to different types of porous media, it has been mainly applied on a relatively small volume of an entire core. Current tools face important limitations for attaining a large representative PN of heterogeneous domains. There are both (1) technical limits and (2) computational limits:
\begin{itemize}
\item Technical limits: There are technical limits on the size of the core that can be scanned to discern void space well. This is rooted in the trade-off of size and resolution and reflected in both the scanning time and size of produced images (\citealt{bultreys_2015}). For instance, typical sandstones require few microns (about $2$ to $5$ $\mu m$) of resolution in CT images to be used in pore-scale modeling while typical scans are not larger than $1000\textsuperscript{3}-2000\textsuperscript{3}$ voxels with the current micro-CT scanners (\citealt{blunt_2017}).
\item Computational limits: Current PN extraction algorithms are not computationally able to provide a large PN covering all heterogeneities at the core-scale. These algorithms generally scan the entire voxelized pore space and the current codes are written as serial codes. Therefore, their performance is limited in terms of both computational memory and speed that results a limit on the size of their input. For example, current MB network extraction codes cannot attain PNs beyond sizes more than roughly $1500\textsuperscript{3}-2000\textsuperscript{3}$ voxels.
\end{itemize}

There are different studies that have addressed some of the challenges noted above by different strategies including multiscale frameworks and domain decomposition approaches in addition to exploiting growing computational capabilities in using current tools.

One common motivation in conducting multiscale studies comes from the practice of investigation of core-scale images (resolution of about $10-30$ $\mu m$) of different heterogeneous rocks in terms of spatial variability that shows a wide range of heterogeneities. While such images can reveal some details of the rock such as connectivity of larger pores, they cannot be directly used in pore-scale modeling tools due to lack of enough precision and failure in capturing small channels and pores which are in sub-resolution. However, these images are still able to cover a large domain of the core thereby allowing study of the heterogeneity and spatial variation of the rock properties. They can be used along with pore-scale images in a multiscale framework to study rock heterogeneity that has been addressed in some studies.

\cite{chu_2013} developed a multiscale algorithm with the form of heterogeneous multiscale method that couples PN models with continuum models to predict pressure and track the macroscopic front, although their work was limited to simpler PN models and did not address pore-scale heterogneities. In a PN study, \cite{jiang_2013} combined coarse-scale and fine-scale pore elements to develop a multiscale PN by their stochastic generator for vuggy carbonate rocks. Although this work could cover three-level PNs from different resolutions, it mainly addressed micropores and its accuracy in prediction of two-phase flow properties was not assessed. This approach was continued by \cite{pak_2016} later where they developed an effective PN integration process to select number and length-scale of input PNs by utilizing experimental pore-throat size distribution. While this approach could combine information from different resolutions to construct a multiscale PNs, it required imaging and PN extraction at all scales and its capability in relative permeability prediction was not studied either. In another multiscale work, \cite{bultreys_2015} developed a dual PN model that incorporates microporosity via adding micro-links to traditional PN models to address the small-scale heterogeneities. While their approach results a great number of pore elements in the constructed PN, it was mainly designed for modeling micropores of heterogeneous samples.

Some other studies have used domain decomposition to deal with the challenges of modeling in large domain problems. \cite{balhoff_2007} used a domain decomposition approach using mortar spaces in which a pore-scale model can be coupled with an adjacent model (pore-scale or continuum) by a 2D finite-element space (\citealt{balhoff_2007}); \citealt{balhoff_2008}). The objective of using mortar coupling was to model a more realistic boundary conditions for different neighboring pore-scale models rather than to provide a large representative volume that includes heterogeneities. In another mortar space study, \cite{sun_2012} used a domain decomposition method to upscale absolute permeability on heterogeneous PN models. Also, \cite{mehmani_2014} used a hybrid method for parallel modeling of linear and nonlinear flow across scales in large domains that enhanced the computational efficiency of mortar domain decomposition. These studies successfully bridged the pore-scale to continuum-scale but were mainly validated based on single-phase flow properties and did not involve core-scale heterogeneities. In a different domain decomposition study, \citet{rabbani_2019} connected extracted PNs of subdomains by generating pore-throats between subdomains with different decomposition strategies. While their approach was successful in reducing computational demands of PN extraction, it was mainly focused on PN properties and absolute permeability of homogeneous samples. In a large domain study, \cite{da_2019} used dual grid domain decomposition on a large rock sample with the aid of high performance computing but the work was limited to absolute permeability evaluation.

In a pore-to-core PN modeling work, \cite{aghaei_2015} studied a long Berea sandstone core to attain a large PN and model capillary and viscosity effects with the aid of heavy parallelization based on domain decomposition scheme. While this work was notable in terms of the accuracy of prediction of two-phase flow properties and the size of studied sample, the type of rock was relatively homogeneous and the process of connecting sample pieces were specific to that very studied sample. In a recent pore-to-core PN modeling study, \cite{zahasky_2019} characterized centimeter-scale Bentheimer sandstone cores by studying heterogeneity in capillary pressure and relative permeability with the aid of an extensive multiple resolution imaging. In addition, \cite{jackson_2019} used multiscale experimental and modeling approaches to study two large Bentheimer sandstone cores. Although both works were notable in terms of addressing heterogeneity and size of the domain, their workflow demands extremely high storage and computational resources.

Therefore, there is still a need for pore-to-core upscaling approaches that can accurately represent the 3D complex pore structure and heterogeneity of real media with efficient computational demands and incorporate information from multiple scales. Herein, we introduce a novel pore-network stitching method (PNSM) to provide large-enough representative PN for a core that encompasses a larger scale of heterogeneities than is possible using conventional PN modeling and other pore-scale modeling approaches while it is still computationally efficient. Some highlights of this method were originally presented by us in \cite{kohanpur_2018_agu} and \cite{kohanpur_2019_agu}.

The organization of the rest of this paper is as follows. Section \ref{method} will explain the overall workflow, steps, and  tools used in the PNSM. In Section \ref{validation}, the main steps of this method, layered stitching and volumetric stitching, are validated for computing flow properties on different types of rock samples. The main application of the PNSM for larger domains is presented in Section \ref{results} where spatially separated PNs are connected together and a large heterogeneous sample is modeled from the information of its signature parts. We compare single-phase and two-phase flow properties of a stitched PN with properties from a defined reference PN for each study.

\section{Method}
\label{method}
\begin{figure}[!htbp]
\centering
\includegraphics[width=0.6\textwidth]{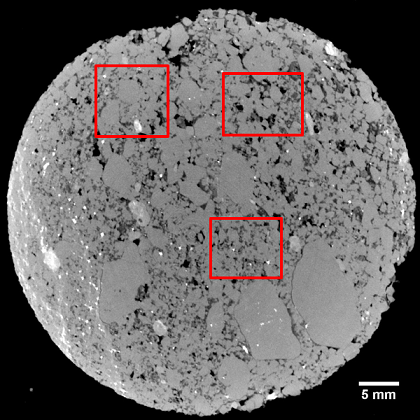}
\caption{A coarse-scale scan of a Mt. Simon sandstone sample at a depth of 7034 feet from the Illinois Basin Decatur Project.}
\label{fig_method-core}
\end{figure}

As mentioned earlier, the motivation for developing the PNSM comes from investigation of core-scale images of heterogeneous rocks that can represent an entire cross section of a core along its length. For example, Fig. \ref{fig_method-core} shows a cross section of a heterogeneous Mt. Simon sandstone core with a diameter of 2 inches at a depth of 7034 feet. The formation is located at verification well number 2 of Illinois Basin-Decatur Project (\citealt{finley_2014}) where Illinois State Geological Survey carried out a pilot CO\textsubscript{2} injection study. The core plug from the formation was scanned by industrial CT scanning at the National Energy Technology Laboratory (NETL) of the U.S. Department of Energy. The marked red boxes in Fig. \ref{fig_method-core} are some distinct regions or heterogeneities of the core that characterize the overall variability. In general, these regions can be found in 3D images of the core at different depths but here we use this 2D image to explain steps taken in the workflow more intuitively.

The first step is to examine the entire coarse resolution scans of the core to find unique size and shape of solid grains and local pore structure in different locations and identify heterogeneities. Assessment and integration of multiple scales of 2D and 3D micro-CT images can be used to investigate geometry, connectivity, and heterogeneity of pore space (\citealt{long_2013}). This procedure might be tedious and subjective but it is an important step in the workflow. It can be doubled-checked by applying a fast evaluation of pore size distribution on multiple 2D images (\citealt{munch_2008}) from different selected depths where larger pores can still be identified in coarse resolution images and most heterogeneities can be captured. The outcome of this analysis is identification of selected locations of the core that are unique in terms of pore size and structure that are defined as signature parts, as shown with red boxes in Fig. \ref{fig_method-core}. The hypothesis is that these signature parts, which might or might not be adjacent, are sufficient to represent the heterogeneity of the core. 

\begin{figure}[!htbp]
\centering
\includegraphics[width=0.95\textwidth]{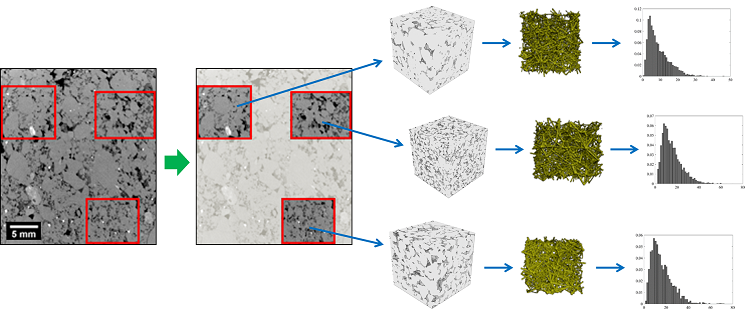}
\caption{Signature parts of the core are selected and their corresponding micro-CT scans, extracted pore-networks, and statistics are derived.}
\label{fig_method-steps-1}
\end{figure}

As depicted in Fig. \ref{fig_method-steps-1}, the next step after identifying signature parts is to obtain their fine resolution scans, 3D micro-CT images, in order to characterize their pore structure. These micro-CT images are used as inputs to a PN extraction tool that obtains the PNs for each signature volume. These extracted PNs also include key statistics of the pore space such as pore size distribution and connectivity. In this workflow, we use the PN extraction code based on MB algorithm from \cite{dong_2009}. The algorithm was originally introduced by \cite{silin_2006} where the entire 3D voxelized pore space is searched to find the largest possible spheres, and subsequently, was extended and modified in later works such as \cite{alkharusi_2007}, \cite{dong_2009}, and \cite{raeini_2017}.

The input of this algorithm is the voxelized binarized (solid as one and pore as zero) geometry of the rock where all the zero voxels are scanned. The largest possible voxelized sphere for each pore voxel is determined and taken as the MB. In practice, most of MBs are removed since they are completely inside the larger MBs. The resulting MBs are sorted and clustered based on their volume that helps identifying the ancestor MBs and a chain of the MBs that are from one ancestor to another one. Each chain is segmented as a configuration of two pore-bodies and their connecting pore-throat (\citealt{dong_2009}). The extracted PN is defined when all pore-bodies and pore-throats are identified in the entire volume of zero voxels.

Finally, the algorithm assigns indices to pore-bodies and pore-throats separately that is used for storing topological information i.e. connections between pore elements. In addition, geometrical information of pore elements are stored including the location, radius, volume, length, total length, and shape factor. The shape factor is a metric of irregularities of the pore space in pore elements and defined as $G = VL/A^2$, where $A$ is surface area, $V$ is volume of the voxelized element, and $L$ is twice the distance between the center of the ancestor MB and the farthest voxel in that MB (\citealt{dong_2009}). The shape factor is also a key quantity that helps assigning familiar geometries (such as circles, squares, and triangles) to the cross sections of the pore elements to be used in the PN flow models (\citealt{patzek_2000}). We use statistics of geometrical and topological information of pore elements in the signature PNs in this workflow.

\begin{figure}[!htbp]
\centering
\includegraphics[width=0.95\textwidth]{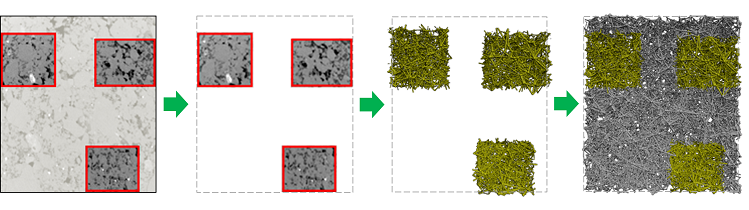}
\caption{Signature pore-networks of the core are used to generate new pore elements in the empty region of the domain and construct a large stitched pore-network.}
\label{fig_method-steps-2}
\end{figure}

Since the space between signature PNs in the core can be relatively large and cannot be directly extracted, the next important step of the workflow is to fill the space between signature PNs by new defined pore elements. We accomplish this by using statistics of the signature PNs and a stochastic algorithm to generate new PNs in the empty regions of the domain. We then develop a new PN stitching algorithm to link adjacent PNs. The final result of the workflow are multiple realizations of a large PN for the entire core volume that represents the heterogeneity identified in the different signature parts. Fig. \ref{fig_method-steps-2} summarizes the path from signature parts of the core to generate new pore elements in the empty region of the domain and construct a large PN as the output that includes both extracted and generated pore elements.

We explain the two main tasks, namely, defining a layer of pore elements linking neighboring PNs and generating an entire PN in empty regions of the domain in Sections \ref{sec_meth_lay} and \ref{sec_meth_vol}, respectively.  

\subsection{Layered stitching}
\label{sec_meth_lay}
\begin{figure}[!htbp]
\centering
\includegraphics[width=0.70\textwidth]{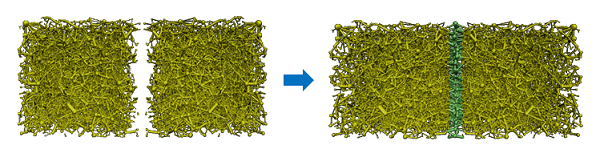}
\caption{Layered stitching of two adjacent pore-networks by generating a layer of pore elements.}
\label{fig_method-stitching_lay}
\end{figure}

The goal in layered stitching is to connect two neighboring PNs by a layer of pore elements that can be in the direction of flow (longitudinal) or transverse directions to the flow (lateral). A layer of pore elements signifies that the thickness of layer is in the range of average pore-throat length as opposed to an entire volume of PN. This step is functionally similar to domain decomposition approach using mortar spaces, since the purpose is to connect two adjacent PNs. We will explain the details by considering longitudinal stitching of left and right PNs together, as shown Fig. \ref{fig_method-stitching_lay}, where the flow direction is from left to right in the flow model.

Initially, the algorithm reads in all information of the left and right PNs, represented in yellow in Fig. \ref{fig_method-stitching_lay}, including: index, location, radius, length, shape factor, volume, connectivity, and inlet and outlet status of all pore-bodies and pore-throats. The next step is to remove outlet pore-throats of left PN and inlet pore-throats of right PN based on the flow direction assumption. This affects the inlet and outlet status of corresponding pore-bodies in those locations. The stitching layer includes interconnected pore-bodies and pore-throats which are connected to the left and right PNs, as represented in green in Fig. \ref{fig_method-stitching_lay}. The thickness of the stitching layer is based on the summation of average pore-throat length of both PNs. Pore-bodies in the stitching layer are defined first where their centers are initially located on a regular 2D lattice pattern proportional to dimensions of the cross section and followed by a random perturbation in 3D space. To assign the number of pore-bodies in the stitching layer, density of pore-bodies (number of elements per the box volume) for each PN is calculated and their arithmetic mean is computed. Then, by multiplying this average density by the thickness and area of the the stitching layer, one can obtain the corresponding number of pore-bodies in the stitching layer. In this way, it is guaranteed that no two pore-bodies intersect and they are not too close to each other, otherwise it could yield extremely short pore-throats in the vicinity of the stitching layer which is not realistic and may affect the resulting flow through the layer.

In the next step, the generated pore-bodies of the stitching layer need to get connected to each other and nearby pore-bodies in both PNs i.e. generating pore-throats of the stitching layer. To do so, initially all possible connections up to average pore-throat length plus two standard deviations are generated. Then, we remove some connections randomly to achieve a determined number of pore-throats in the stitching layer which is the arithmetic mean of density of pore-throats from both PNs. However, it is also important to investigate the average connection number of the final stitched PN and compare it with the arithmetic mean of the original PNs. Therefore, a further step of adding or removing some pore-throats in the stitching layer is applied to get the average connection number within a threshold of the arithmetic mean of left and right PNs.

After defining number and location of pore elements and their connections in the stitching layer, their corresponding radius, shape factor, and volume need to be determined. To do so, we add distributions of left and right PNs for each property by combining their histograms and fitting a proper distribution function such as normal, log-normal, and Weibull distributions. These distributions are also used in other PN modeling works such as \cite{valvatne_2003}. While we can choose different combinations of these distributions and evaluate the best one based on final resulting flow properties, we assign default distributions for each property in the workflow coming from studying a wide range extracted PNs from different types of rock. Based on our analysis, the Weibull distribution is a better candidate for radius and volume and normal distribution for shape factor of pore elements in most of studied cases, so these are the default distributions in our developed code.

The explained procedure is applicable when the defined layer of pore elements is normal to the main flow direction. On the other hand, a similar strategy can be applied when the two neighboring PNs are stitched together in the lateral direction which is parallel to the main flow direction. In such case, we define the stitching layer with just generated pore-throats coming from average statistics of both PNs and no pore-bodies are generated in the lateral direction. In this way, we mark all pore-bodies of both PNs in the vicinity of stitching layer (e.g., within the $10\%$ of lateral length) and generate all possible connections. Then, we apply a similar removal approach to achieve a determined number of pore-throats in the stitching layer based on considering arithmetic mean of density of pore-throats and average connection number of both PNs. The radius, shape factor, and volume of these generated pore-throats are also determined by fitting a proper distribution function as discussed for longitudinal layered stitching.

Thus, the size and connectivity of the layer of pore elements in longitudinal and lateral directions are influenced by the statistics of the two adjacent PNs. We implement the stitching layer as a deterministic approach to reduce complexity, albeit a smooth connection between adjacent PNs should be achieved and tested in terms of resulting flow properties. In Sections \ref{sec_val_laylong} and \ref{sec_val_laylat}, both longitudinal and lateral stitching procedures are validated on several samples, respectively.

\subsection{Volumetric stitching}
\label{sec_meth_vol}
\begin{figure}[!htbp]
\centering
\includegraphics[width=0.95\textwidth]{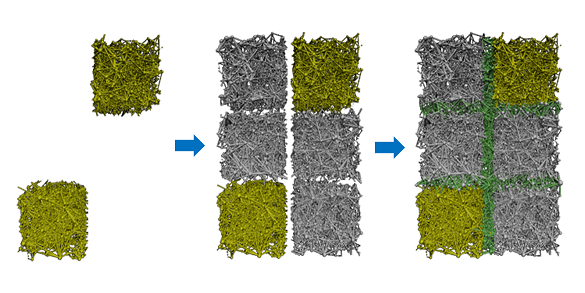}
\caption{Volumetric stitching of two distant pore-networks by generating pore elements in the empty region and linking all pore-networks by layered stitching.}
\label{fig_method-stitching_vol}
\end{figure}

If the signature PNs are not neighbors in the domain, which is usually the case, layered stitching would not be sufficient to connect them and another way of generating pore elements in the empty volume of the domain is needed that we call volumetric stitching. We use PN generators to have similar box-size PNs and place them in empty spaces including the vicinity of signature PNs as shown in Fig. \ref{fig_method-stitching_vol}. This approach allows using available PN generators in the literature, such as the stochastic network generator developed by \cite{idowu_2010}, which we have used in this workflow.

This generator utilizes detailed statistics of pore elements as input data and results a generated PN in the same format as an extracted PN from MB algorithm. To be more specific, the required input data are distributions of pore-body radius and volume, connection number, pore-throat radius and volume, pore-throat length, and pore-throat total length along with their associated indices. The generator assigns locations of all pore-bodies randomly first and, then it reconciles pore-throat lengths and connection numbers to generate pore-throats between pore-bodies. Thus, the resulting locations of pore-elements and statistics of pore-throats are stochastic and varies across realizations. As a result, the output is a stochastic generated PN and sufficient number of realizations must be generated for any analysis. More details can be found in \cite{idowu_2010}.

\begin{figure}[!htbp]
\centering
\includegraphics[width=0.80\textwidth]{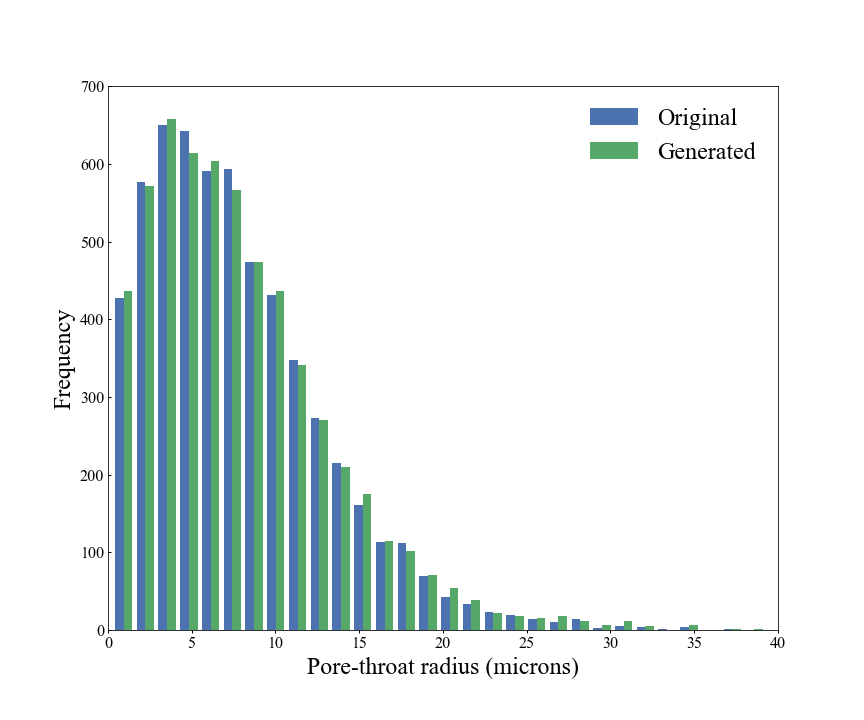}
\caption{Comparison of pore-throat size distribution of a generated pore-network with its original pore-network from a Mt. Simon sandstone sample.}
\label{fig_gen-dist}
\end{figure}

We tested this generator on various PNs coming from different rocks and observed that it sometimes fails to keep the connectivity of source statistics in the generated PN, and therefore, underestimates permeability. To fix this issue, we have added a modifier after generating PNs to check the weighted average connection number (a metric that describes size and connectivity of PN together) and adjust the pore-throat size distribution. In this way, some of the generated realizations are rejected if their weighted average connection number is not within the $10\%$ of the weighted average connection number of source data. Fig. \ref{fig_gen-dist} shows an example of pore-throat size distribution of an accepted realization of generated PN compared with the original PN from a Mt. Simon sandstone sample that its statistics are used as the source data.

The explained selective approach in PN generation leads to better realizations of generated PNs for filling the empty regions of the domain. The statistics of the generated PNs should be derived from those of nearby signature parts. One such approach would be to use parameters (e.g., mean, variance) of derived fitting distributions. We accomplish this by concatenating the pore element lists (pore-body radius and volume, connection number, pore-throat radius and volume, pore-throat length, and pore-throat total length) for all signature PNs with a weight for each based on their relative center-to-center distance to the generated PN i.e. the closer to a signature PN, the more it is influenced by its statistics. The advantage of this way is twofold. First, we can easily honor the intrinsic correlation among geometrical properties of a pore element such as radius, length, volume, etc. by picking them as a set as opposed to considering separate distributions and then matching correlations. This leads to having more realistic pore elements. Second, we can relate the weighted average to all signature PNs of the entire domain simultaneously and thus, the footprint of their statistics can be found in every single generated PN of the domain.

Once all empty spaces are filled with generated PNs, we use the layered stitching in longitudinal and lateral directions to connect all PNs together as shown in Fig. \ref{fig_method-stitching_vol}. Therefore, the stochastic PN generation is coupled with layered stitching to create a large stitching volume to implement volumetric stitching and connect signature PNs that might be in distant locations in the core. It should be emphasized that the combination of stochastic PN generation and deterministic layered stitching results a stochastic process for volumetric stitching.

An important challenge in the described workflow that happens in most of steps is to sort and relate indices of pore-bodies and pore-throats. This is because each of the signature and stitching PNs has its own order indices of pore-bodies and pore-throats initially but they need to to be modified and reordered by any removal and addition of pore elements. The goal of having a final unified PN requires having one single set of indices for all elements in the final stitched PN. Moreover, the connection number of pore-bodies can change with any removal and addition of pore-throats and pore-bodies; that requires updating and matching indices in the connectivity information of the final unified PN as well.

In summary, the introduced steps together form a workflow to deal with large domain problems that starts with identifying signature parts and having their 3D images. PN extraction is used, and the resulting statistics are used in layered stitching and PN generation to fill empty spaces of the domain. The outcome is that all signature and generated pore elements interconnected in a large unified PN that can be used in flow simulation and other PN applications.

\section{Validation}
\label{validation}
In order to validate the introduced PNSM and its specific steps of stitching explained in Section \ref{method}, various test cases of flow simulations are carried out on different types of rocks. For this validation, six cubic rock samples (four sandstones and two carbonates) are chosen from different studies in the literature; the samples are listed in Table \ref{tab_samples} with their size, image resolution, porosity, and a chosen label.

\begin{table}[!htbp]
\centering
\caption{Selected samples for validating steps of the pore-network stitching method}
\begin{tabular}{lllllll}
\hline
Sample                & Label & Porosity & \begin{tabular}[c]{@{}l@{}}Resolution\\   ($\mu m$)\end{tabular} & \begin{tabular}[c]{@{}l@{}}Length\\   (pixel)\end{tabular} & \begin{tabular}[c]{@{}l@{}}Size\\   ($mm^3$)\end{tabular} & Study     \\
\hline
Berea sandstone         & BR & 0.208 & 3.20 & 400 & 1.28\textsuperscript{3} & \cite{jiang_2014}     \\
Bentheimer sandstone    & BN & 0.188 & 3.18 & 500 & 1.59\textsuperscript{3} & \cite{herring_2016}   \\
Mt. Simon sandstone     & MH & 0.261 & 2.80 & 500 & 1.40\textsuperscript{3} & \cite{kohanpur_2020}  \\
Mt. Simon sandstone     & ML & 0.097 & 1.95 & 800 & 1.56\textsuperscript{3} & \cite{tahmasebi_2017} \\
C1 carbonate            & C1 & 0.233 & 2.85 & 400 & 1.14\textsuperscript{3} & \cite{dong_2009}      \\
C2 carbonate            & C2 & 0.168 & 5.35 & 400 & 2.14\textsuperscript{3} & \cite{dong_2009}      \\
\hline
\end{tabular}
\label{tab_samples}
\end{table}

We carry out single-phase and two-phase flow simulations for different steps of the PNSM on these six samples. As our original motivation was to study geological storage of CO\textsubscript{2}, the CO\textsubscript{2}-brine flow system (see Table \ref{tab_props} for the properties) is used for the two-phase flow simulations, and we use a quasi-static PN flow solver (\citealt{valvatne_2004}) to study its capillary-dominated physics. In each validation study, the stitched PN is compared with a similar-size reference PN that is constructed by connecting 3D image of the original sample to its mirrored geometry along the intended stitching direction.

\begin{table}[!htbp]
\centering
\caption{Properties of CO\textsubscript{2}-brine flow system in pore-network two-phase flow simulations}
\begin{tabular}{lll}
\hline
Properties          & Value & Unit    \\
\hline
Interfacial tension & 30    & \( mN/m \)   \\
Contact angle       & 25    & \( ^{\circ} \) \\
Brine density       & 280   & \( kg/m3 \)   \\
CO\textsubscript{2} density         & 992   & \( kg/m3 \)  \\
Brine viscosity     & 0.550 & \( cp \)      \\
CO\textsubscript{2} viscosity       & 0.023 & \( cp \)  \\
\hline
\end{tabular}
\label{tab_props}
\end{table}

\begin{figure}[!htbp]
\centering
\includegraphics[width=0.80\textwidth]{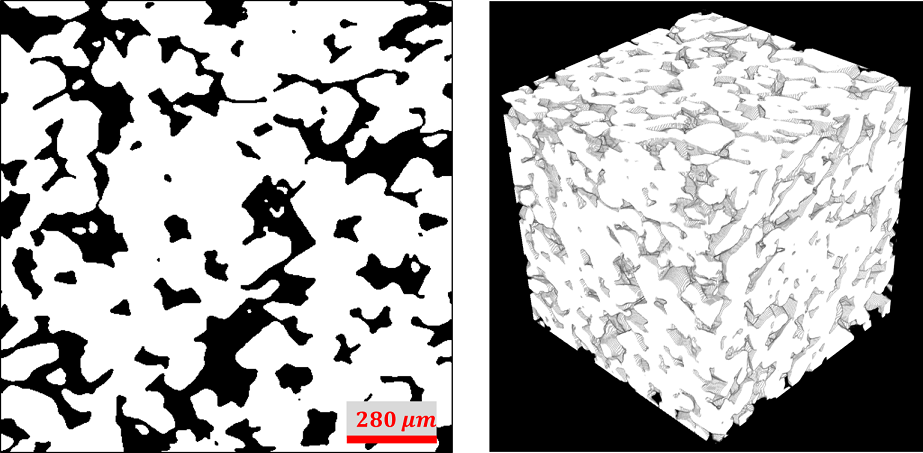}
\caption{The Mt. Simon sandstone sample: (left) the segmented image of the first slice of the stack and (right) the 3D reconstructed geometry of the rock.}
\label{fig_sample-mh}
\end{figure}

For single-phase flow properties, we report absolute permeability of all six samples here. For two-phase properties, we just report the results from the Mt. Simon sample (MH) here which its raw and segmented 3D images are available at \cite{data_mtsimon}. This sample has a fairly high porosity and is more heterogeneous with respect to typical well-studied sandstones in the literature such as Berea and Bentheimer. Since the sample is from the reservoir that has  been used in a pilot CO\textsubscript{2} sequestration project in Illinois (\citealt{finley_2014}), its CO\textsubscript{2}-brine flow properties are extensively studied recently by different numerical pore-scale models (\citealt{kohanpur_2020}). We use subsample S2 from that study, labeled MH in Table \ref{tab_samples} and shown in Fig. \ref{fig_sample-mh}, to validate each step of the PNSM. The full results on other samples of Table \ref{tab_samples} are available in Supplementary Material of this paper.

\subsection{Longitudinal layered stitching}
\label{sec_val_laylong}
\begin{figure}[!htbp]
\centering
\includegraphics[width=0.80\textwidth]{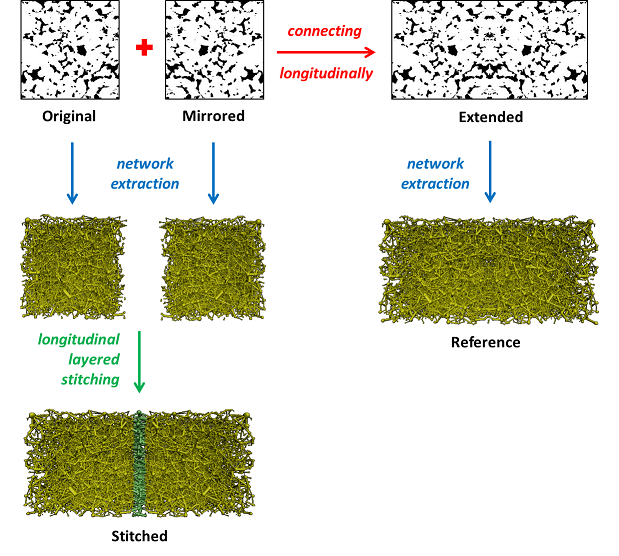}
\caption{Longitudinal layered stitching of a Mt. Simon sandstone sample to its mirrored geometry and its corresponding reference sample.}
\label{fig_val-laylong_stitching}
\end{figure}

In order to validate the layered stitching in longitudinal direction (along the main flow direction), the original sample is considered as the left sample and its 3D image is mirrored longitudinally to obtain the right sample. Connecting left and right samples in the voxelized domain forms the extended sample as shown in Fig. \ref{fig_val-laylong_stitching}. By applying PN extraction algorithm on 3D images of these samples, one can get left, right, and extended PNs as depicted in Fig. \ref{fig_val-laylong_stitching}.

While the extended PN is used as the reference PN, left and right PNs are stitched together using the longitudinal layered stitching described in Section \ref{sec_meth_lay}. The stitched and reference PNs are then used as input PNs of the quasi-static PN flow solver (\citealt{valvatne_2004}) with the same input properties to be compared.

\begin{figure}[!htbp]
\centering
\includegraphics[width=0.70\textwidth]{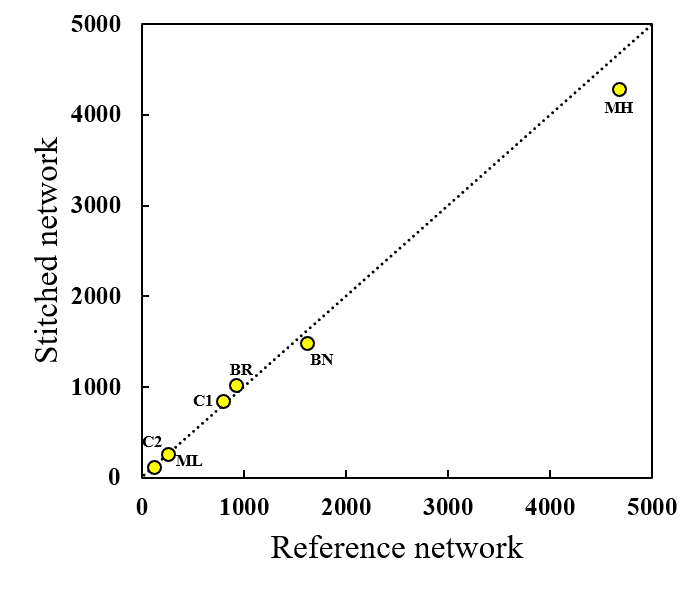}
\caption{Comparison of absolute permeability of reference pore-network with the resulting stitched pore-network via longitudinal layered stitching for all six selected samples.}
\label{fig_val-laylong_k}
\end{figure}

\begin{figure}[!htbp]
\centering
\includegraphics[width=0.80\textwidth]{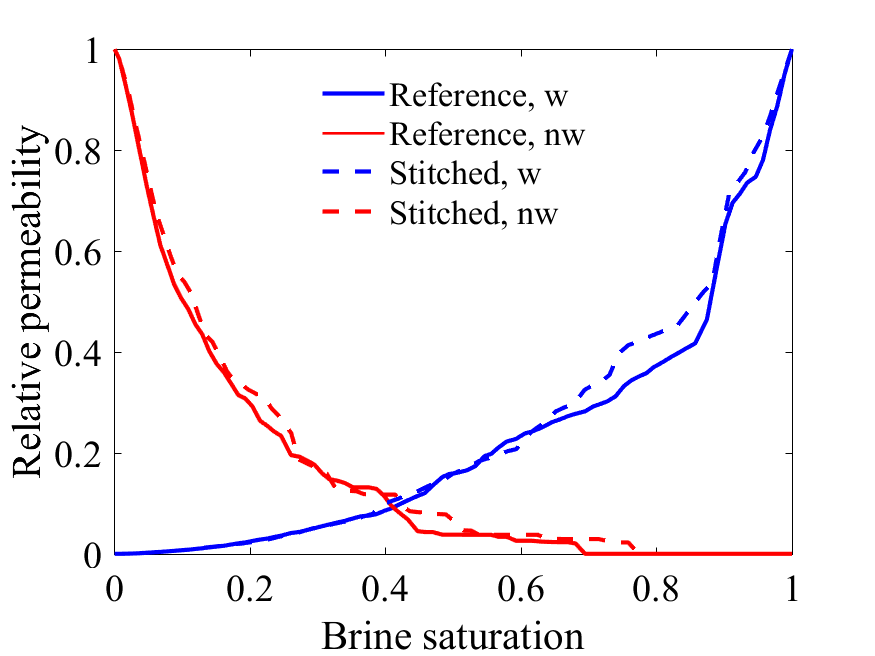}
\caption{Comparison of relative permeability curves of the stitched pore-network from the Mt. Simon sandstone sample (MH) via longitudinal layered stitching with its reference pore-network.}
\label{fig_val-laylong_mh-kr}
\end{figure}

The predicted absolute permeability of all samples are compared in Fig. \ref{fig_val-laylong_k} where horizontal and vertical axes refers to predicted absolute permeability of the reference and stitched PNs, respectively. All samples have results that are quite close to the diagonal line, thus providing confidence in the longitudinal stitching algorithm across a variety of rock types. Regarding two-phase flow properties, Fig. \ref{fig_val-laylong_mh-kr} shows excellent agreement for drainage relative permeability curves between the stitched and reference PNs of sample MH.

\subsection{Lateral layered stitching}
\label{sec_val_laylat}
\begin{figure}[!htbp]
\centering
\includegraphics[width=0.80\textwidth]{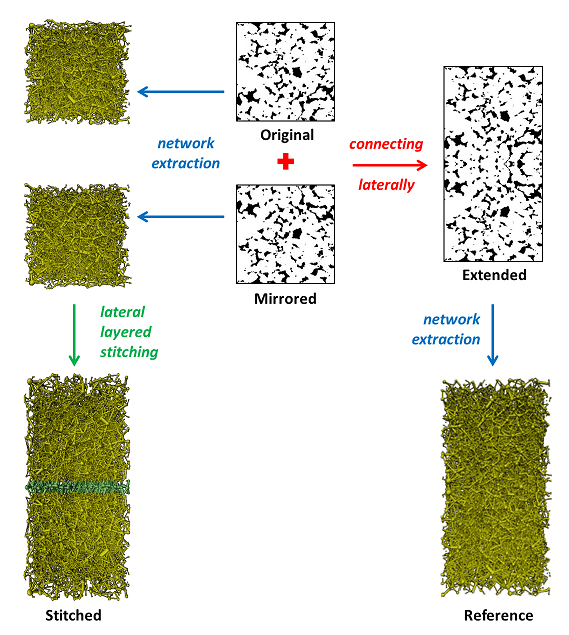}
\caption{Lateral layered stitching of a Mt. Simon sandstone sample to its mirrored geometry and its corresponding reference sample.}
\label{fig_val-laylat_stitching}
\end{figure}

As explained in Section \ref{method}, stitching in the lateral direction is implemented slightly differently from that in the longitudinal direction, adding only pore-throats between PNs. For this validation, the 3D image of the original sample is mirrored laterally and connected to itself to obtain a laterally extended sample as shown in Fig. \ref{fig_val-laylat_stitching}. The PN extraction algorithm is applied on all these pieces to get original, mirrored, and reference PNs as depicted in Fig. \ref{fig_val-laylat_stitching}.

\begin{figure}[!htbp]
\centering
\includegraphics[width=0.70\textwidth]{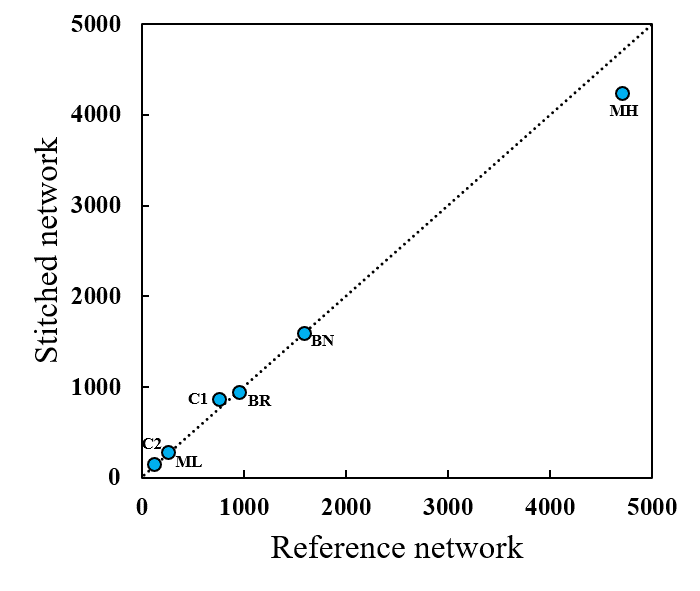}
\caption{Comparison of absolute permeability of reference pore-network with the resulting stitched pore-network via lateral layered stitching for all six selected samples.}
\label{fig_val-laylat_k}
\end{figure}

\begin{figure}[!htbp]
\centering
\includegraphics[width=0.80\textwidth]{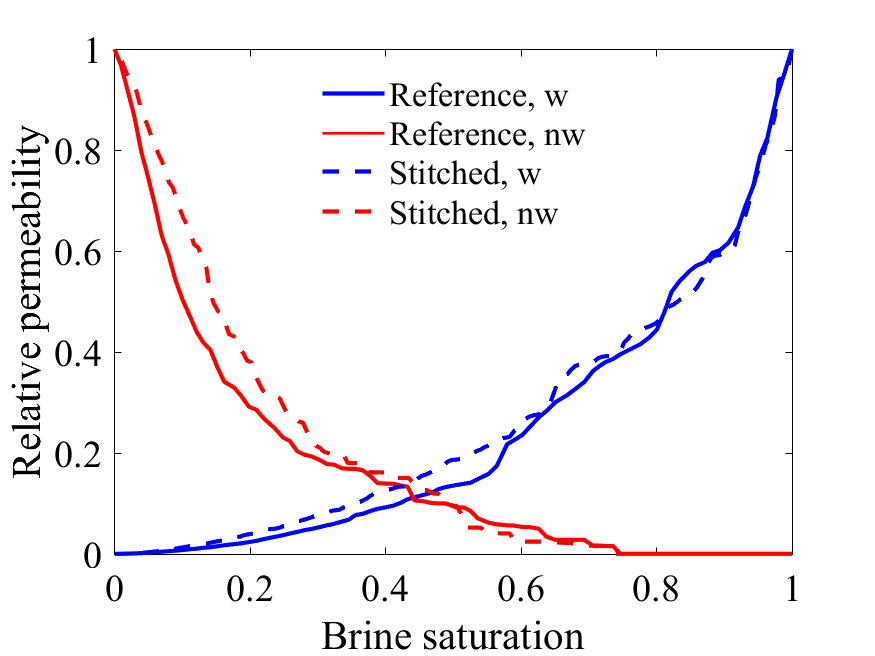}
\caption{Comparison of relative permeability curves of the stitched pore-network from the Mt. Simon sandstone sample (MH) via lateral layered stitching with its reference pore-network.}
\label{fig_val-laylat_mh-kr}
\end{figure}

The original and mirrored PNs are stitched via lateral layered stitching explained in Section \ref{sec_meth_lay} to be compared with the reference PN based on flow properties. The predicted absolute permeability of all samples are compared in Fig. \ref{fig_val-laylat_k}. The results are all close to the diagonal line again, which means a good accuracy of lateral layered stitching across these six samples. Fig. \ref{fig_val-laylat_mh-kr} presents the comparison of drainage relative permeability curves of stitched and reference PNs from sample MH that shows excellent agreement.

Although longitudinal and lateral layered stitching are contributing a small fraction of pore elements in the final stitched PN, the quality of defining new pore elements from the statistics is crucial in terms of connectivity and the resulting flow properties of the stitched PN.

\subsection{Volumetric stitching}
\label{sec_val_vol}
\begin{figure}[!htbp]
\centering
\includegraphics[width=0.80\textwidth]{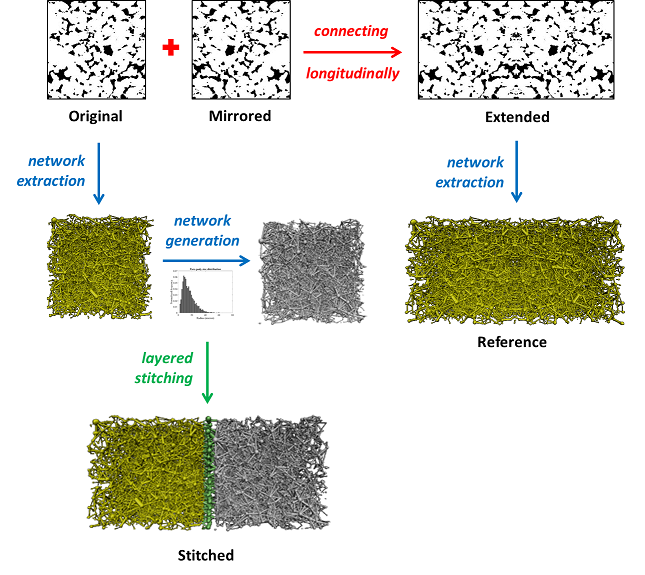}
\caption{Lateral layered stitching of a Mt. Simon sandstone sample to its mirrored geometry and its corresponding reference sample.}
\label{fig_val-vol_stitching}
\end{figure}

After ensuring that layered stitching works smoothly in both longitudinal and lateral directions, it is also necessary to investigate the performance of volumetric stitching where an entirely new PN and stitching layers are generated in the domain. As discussed in Section \ref{method} because the utilized PN generator is a stochastic tool, the volumetric stitching is also stochastic tool, and this needs to be reflected in the evaluation of properties. Fig. \ref{fig_val-vol_stitching} shows the defined problem to evaluate flow properties of the stitched PN via volumetric stitching. The extended geometry along the main flow direction can still be used as the reference PN while the mirrored sample is replaced by a generated PN. Then, longitudinal layered stitching is applied to link the original and generated PNs to obtain one realization of the stitched PN via volumetric stitching.

\begin{figure}[!htbp]
\centering
\includegraphics[width=0.70\textwidth]{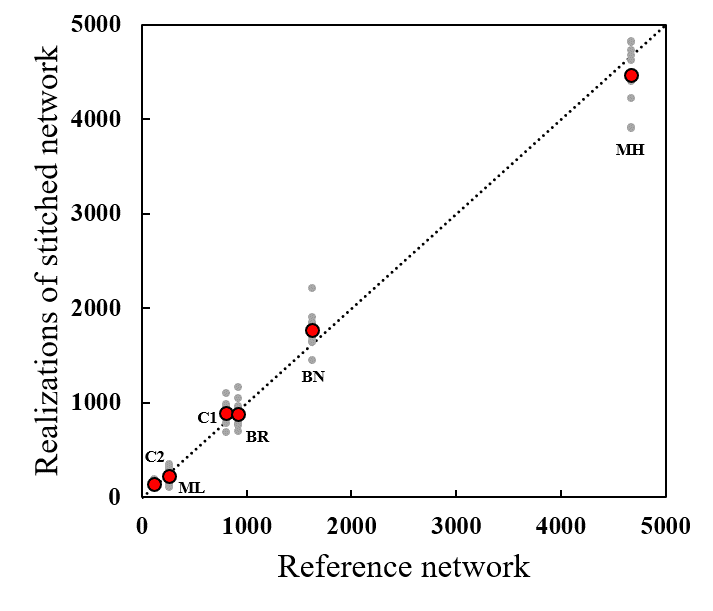}
\caption{Comparison of absolute permeability of reference pore-network with the resulting realizations of stitched pore-network and their mean via volumetric stitching for all six selected samples.}
\label{fig_val-vol_k}
\end{figure}

We have carried out volumetric stitching with 10 realizations on each sample listed in Table \ref{tab_samples}. Fig. \ref{fig_val-vol_k} shows the arithmetic mean and range of the resulting permeability of each sample compared with its reference PN. Since the approach is stochastic, it is possible that some realizations will be outliers in terms of computed flow properties. However, the mean over the 10 realizations is close to the diagonal line, showing good agreement with the reference PN.

\begin{figure}[!htbp]
\centering
\includegraphics[width=0.80\textwidth]{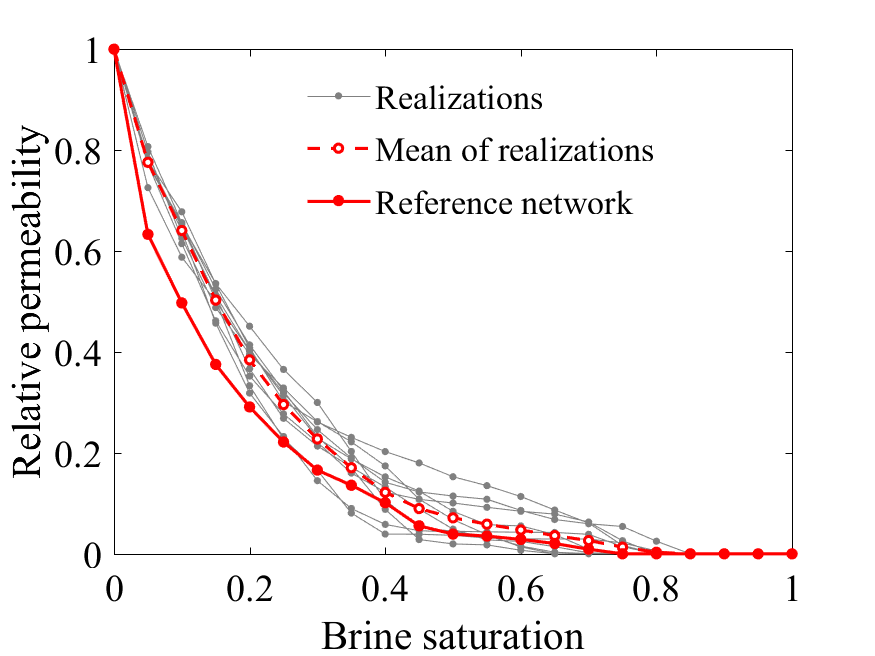}
\caption{Comparison of CO\textsubscript{2} relative permeability of 10 realizations of stitched pore-networks from the Mt. Simon sandstone (MH) sample via volumetric stitching with their reference pore-network.}
\label{fig_val-vol_mh-krn}
\end{figure}

\begin{figure}[!htbp]
\centering
\includegraphics[width=0.80\textwidth]{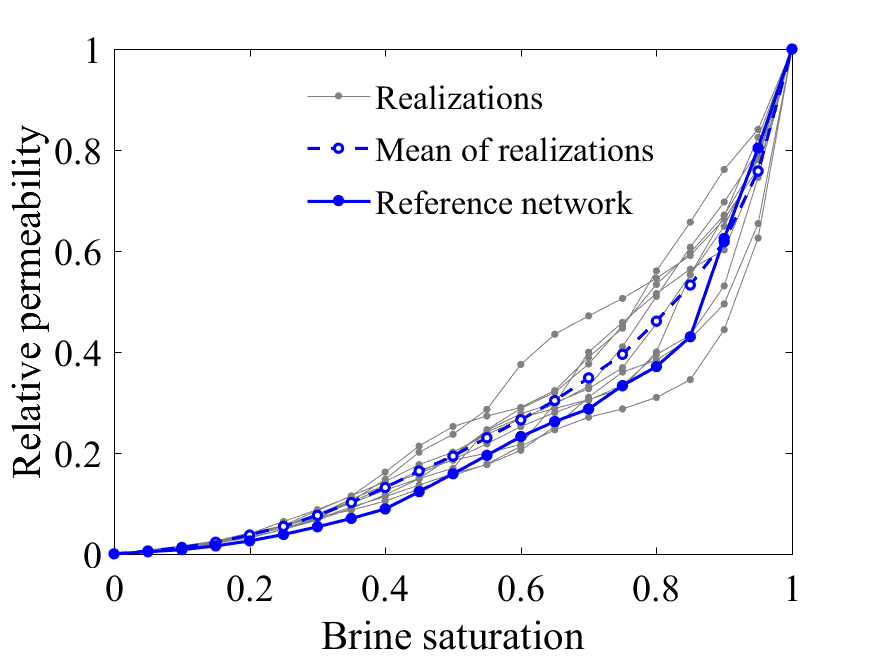}
\caption{Comparison of brine relative permeability of 10 realizations of stitched pore-networks from the Mt. Simon sandstone (MH) sample via volumetric stitching with their reference pore-network.}
\label{fig_val-vol_mh-krw}
\end{figure}

To assess the resulting relative permeability curves, we compare the mean of realizations with the reference PN curves. We have obtained this by taking the arithmetic mean at each saturation point. Figs. \ref{fig_val-vol_mh-krn} and \ref{fig_val-vol_mh-krw} show drainage relative permeability of CO\textsubscript{2} and brine, respectively, for 10 realizations along with their mean compared with the reference curve. The predicted relative permeability of both phases show good agreement between stitched PN and reference PN curves. Relative permeability curves from other samples are reported in Supplementary Material.

It should be emphasized that this stochastic approach is feasible because PN generation and stitching are computationally efficient and quasi-static PN solver is a also fast predictive tool.

\section{Results and Discussion}
\label{results}
In this section, we use the PNSM on two different defined problems with large domains relative to conventional pore-scale studies. The first problem considers two distant samples in the flow direction with different properties where the space between them is filled with new pore elements based on combined statistics of their PNs. The second problem is for a large heterogeneous 3D sample where three small signature parts are used to construct a large representative PN and compute flow properties. The focus of the first problem is to combine different PN statistics to generate an average PN in the empty space in between, which are stitched to original PNs. On the other hand, the focus of the second problem is to deal with different pieces of a large sample and the choice of signature parts as inputs of PNSM. However, the ultimate goal of both problems is to obtain a large unified PN that is used in a flow solver to compute upscaled flow properties. We have defined these two problems in a fashion to be able to have a reference PN to compare predicted properties; however, the PNSM can certainly be applied on larger sizes as well for which it is not feasible to extract a single PN.

\subsection{Connecting distant pore-networks}
\label{sec_res_conn}
\begin{figure}[!htbp]
\centering
\includegraphics[width=0.50\textwidth]{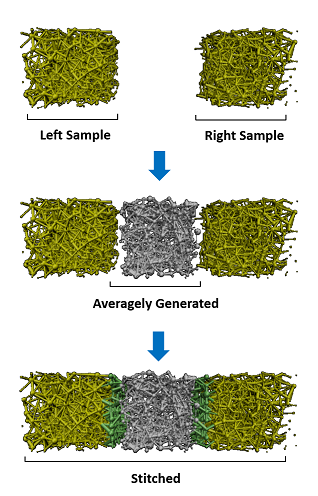}
\caption{Stitching of pore-networks of two distant Mt. Simon sandstone samples by generating an average pore-network in the middle.}
\label{fig_results-distant_stitching}
\end{figure}

In order to investigate the quality of combining statistics of different PNs and a resulting average statistics, a Mt. Simon sandstone core with the size of $1200\times400\times400$ voxels and resolution of $2.80$ $\mu m$ is divided into three pieces (left, middle, and right) of $400^3$ voxels having porosity $0.255$, $0.233$, $0.238$, and absolute permeability $3784$ $mD$, $2935$ $mD$, $2598$ $mD$, respectively from left to right (obtained by PN flow modeling). In the defined problem, the middle piece is removed and the left and right pieces are considered as signature parts of the PNSM. The goal is to combine the information from extracted PN from left and right pieces and to generate a middle PN that connect the two original PNs. Fig. \ref{fig_results-distant_stitching} shows the steps taken to obtain the final stitched PN from the two original PNs at left and right corners of the core.

\begin{figure}[!htbp]
\centering
\includegraphics[width=0.80\textwidth]{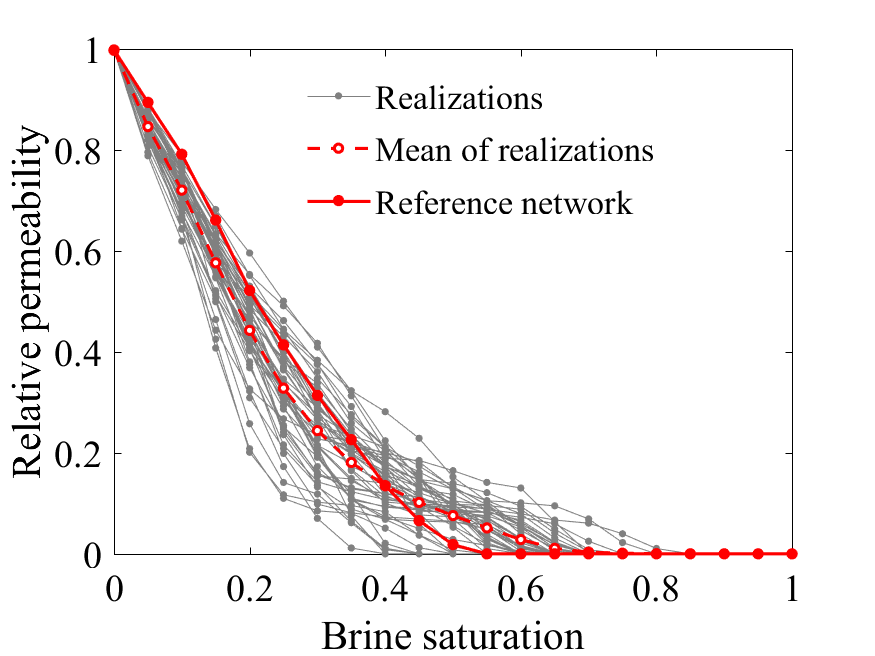}
\caption{Comparison of drainage relative permeability of CO\textsubscript{2} from a long Mt. Simon sample with 50 realizations of stitched pore-networks and their mean.}
\label{fig_results-distant_krn}
\end{figure}

\begin{figure}[!htbp]
\centering
\includegraphics[width=0.80\textwidth]{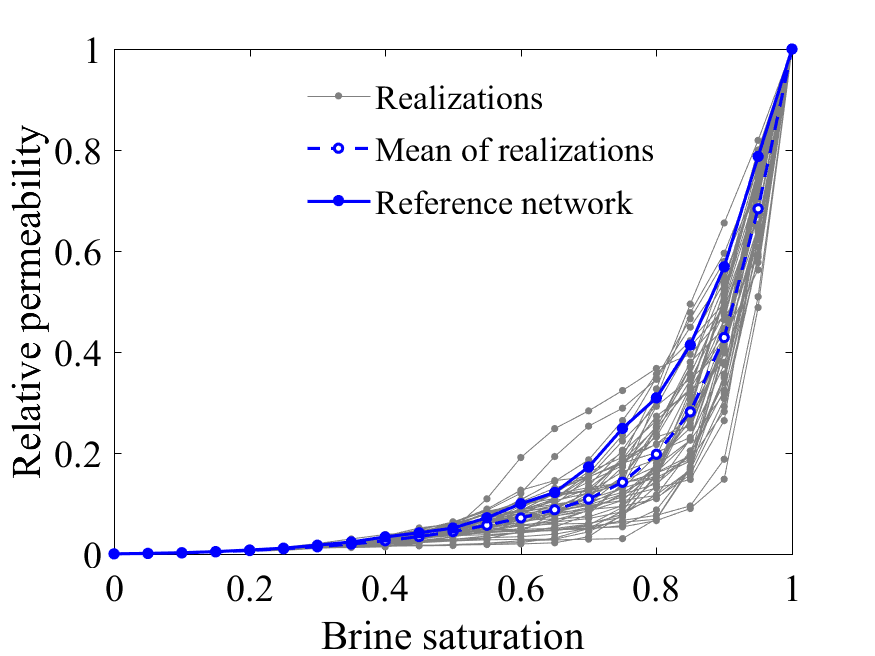}
\caption{Comparison of drainage relative permeability of brine from a long Mt. Simon sample with 50 realizations of stitched pore-networks and their mean.}
\label{fig_results-distant_krw}
\end{figure}

The resulting stitched PN is compared with the extracted PN from the entire core by taking the mean over 50 realizations. Figs. \ref{fig_results-distant_krn} and \ref{fig_results-distant_krw} show drainage relative permeability of CO\textsubscript{2} and brine, respectively, for 50 realizations along with their mean. Regardless of the variation across realizations, the comparison shows good agreement between the mean of the stitched PN curves and the reference PN curve for each phase.

\subsection{Modeling a large heterogeneous sample}
\label{sec_res_het}
In order to apply the developed PNSM on a core-scale problem, we have constructed a large heterogeneous sample ($6.4$ $mm^3$) by extending a Berea sandstone sample in all directions and altering its pore structure in random locations. To do so, we chose random boxes with arbitrary sizes on the 3D image of the extended sample and utilized noise addition or removal functions in Fiji (\citealt{schindelin_2012}) on the pore space of these boxes to expand or shrink the pore structure. The outcome is a wide range of variations in pore structure and a large heterogeneous sample. A stochastic approach is also implemented to quantify the heterogeneity of this sample where 1000 calculation boxes are chosen randomly and the statistics of their weighted average connectivity are calculated. The coefficient of variation of the statistics on the heterogeneous sample is $0.053$ due to the randomized alteration of the pore structure while it is $0.018$ (about one-third) on the homogeneous Berea sandstone and $0.007$ on an equal-size uniform regular PN. The conclusion is that this sample is heterogeneous enough to be used for the actual application of the PNSM.

\begin{figure}[!htbp]
\centering
\includegraphics[width=0.95\textwidth]{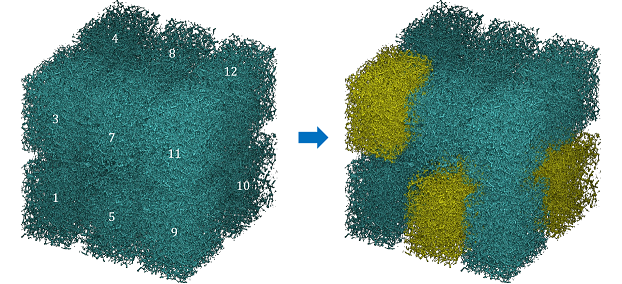}
\caption{A pore-network representation of a constructed heterogeneous sample with its 12 equal-size pieces. Pieces 3, 5, and 10 are selected as signature parts.}
\label{fig_results-het_pieces}
\end{figure}

Fig. \ref{fig_results-het_pieces} shows a PN representation of this sample and how it is divided into 12 equal-size pieces. We labeled them from 1 to 12, starting from bottom left corner as depicted in Fig. \ref{fig_results-het_pieces}, and marked three signature parts (pieces 3, 5, and 10) out of it based on the calculated range of porosity (from $0.152$ to $0.221$) and absolute permeability (from $219$ $mD$ to $1972$ $mD$) across all pieces reported in \ref{tab_het-12}. We have defined a problem where three signature parts are selected to start with and the rest of pieces are removed. The goal of the PNSM on this problem is to fill the empty space with pore elements and stitch them together and to the three signature parts to come up with a large stitched PN covering the entire domain. This PN is compared with a PN extracted from the original heterogeneous sample, as the reference PN, based on flow properties. We look into absolute permeability and drainage relative permeability curves of CO\textsubscript{2}-brine flow across 50 realizations.

\begin{table}[!htbp]
\centering
\caption{Absolute permeability of 12 pieces of the constructed heterogeneous sample}
\begin{tabular}{lll}
\hline
Pieces & Porosity & Permeability ($mD$) \\
\hline
1      & 0.170 & 803    \\
2      & 0.221 & 3387   \\
3      & 0.199 & 1608   \\
4      & 0.182 & 972    \\
5      & 0.170 & 926    \\
6      & 0.201 & 1972   \\
7      & 0.163 & 670    \\
8      & 0.153 & 624    \\
9      & 0.185 & 1075   \\
10     & 0.152 & 219    \\
11     & 0.202 & 1720   \\
12     & 0.175 & 762    \\
\hline
\end{tabular}
\label{tab_het-12}
\end{table}

\begin{figure}[!htbp]
\centering
\includegraphics[width=0.8\textwidth]{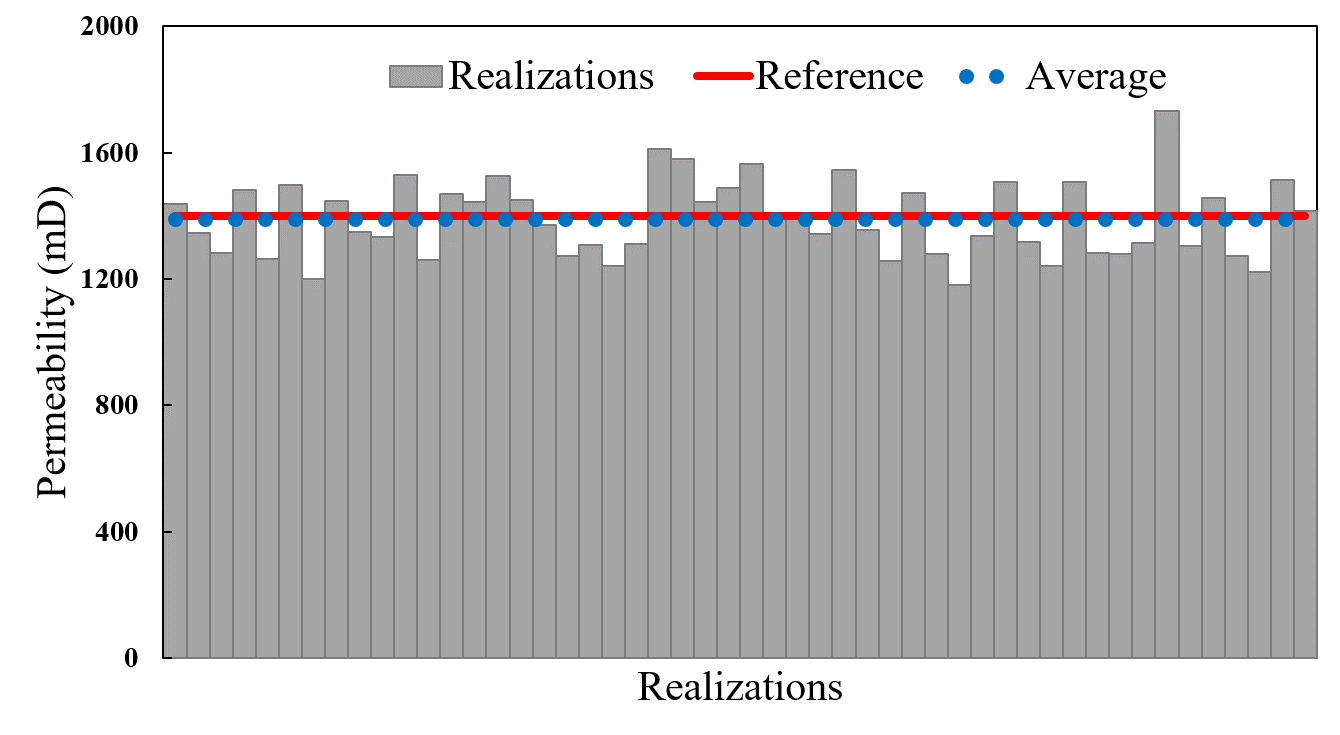}
\caption{Absolute permeability over 50 realizations of stitched pore-networks from the heterogeneous sample.}
\label{fig_results-het_k}
\end{figure}

Fig. \ref{fig_results-het_k} shows the variation of absolute permeability over 50 realizations of stitched PNs having an arithmetic mean equal to $1389$ $mD$, which shows excellent agreement with the reference PN, $1400$ $mD$. This shows that the first three signature pieces could provide enough statistical information about the pore structure to construct a representative PN of the larger sample, and therefore, an accurate prediction of connectivity and permeability is resulted. It is also important to note that we are not seeking a single PN for this purpose but a mean over enough number of realizations due to the stochastic approach in generating pore elements in the initial empty space.

\begin{figure}[!htbp]
\centering
\includegraphics[width=0.8\textwidth]{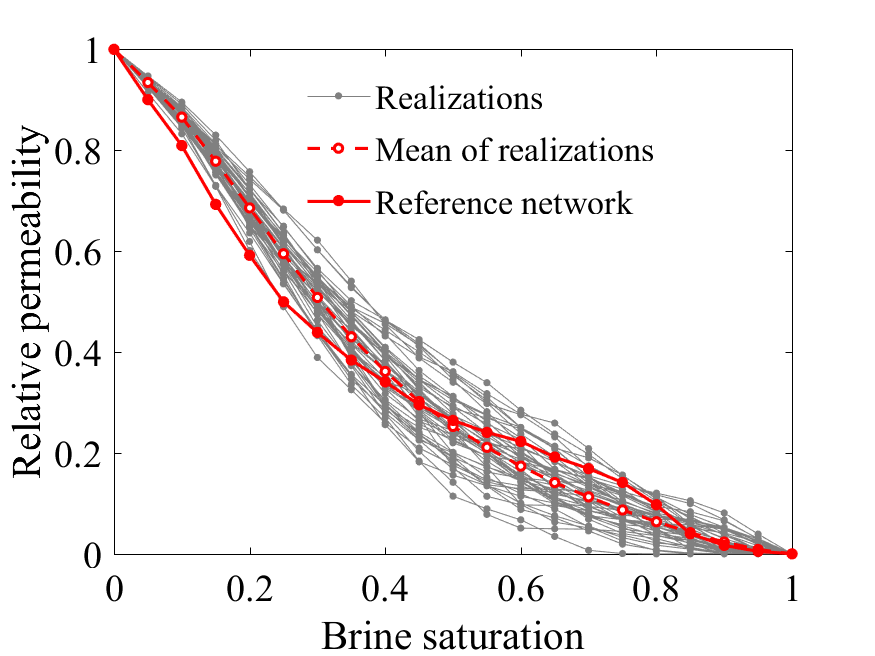}
\caption{Comparison of drainage relative permeability of CO\textsubscript{2} from the large heterogeneous sample with 50 realizations of stitched pore-networks and their mean.}
\label{fig_results-het_krn}
\end{figure}

\begin{figure}[!htbp]
\centering
\includegraphics[width=0.8\textwidth]{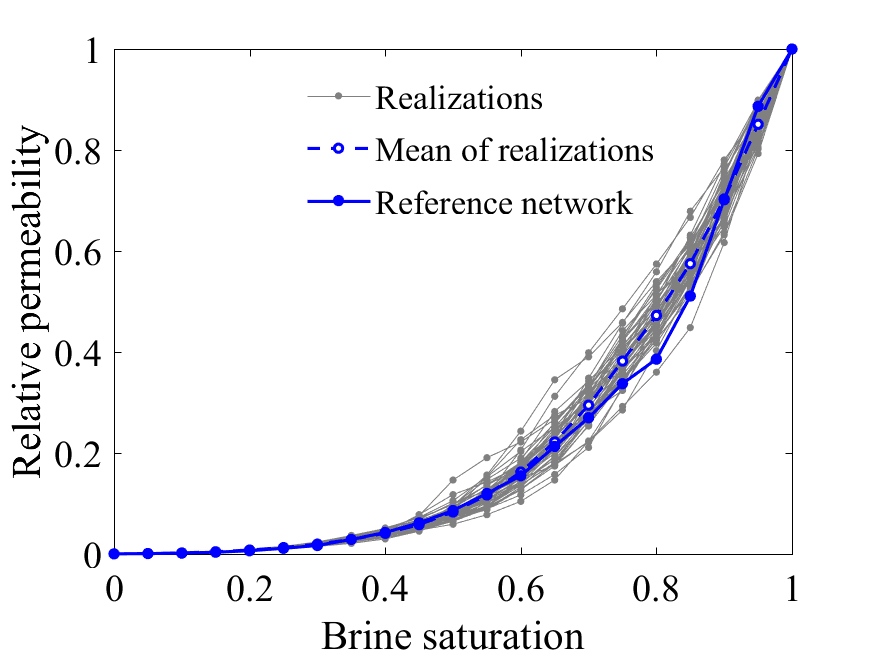}
\caption{Comparison of drainage relative permeability of brine from the large heterogeneous sample with 50 realizations of stitched pore-networks and their mean.}
\label{fig_results-het_krw}
\end{figure}

Figs. \ref{fig_results-het_krn} and \ref{fig_results-het_krw} show the comparative drainage relative permeability results between reference PN and realizations of stitched PN and their mean for CO\textsubscript{2} and brine, respectively, as a function of brine saturation. The agreement for both phases is quite good when we compare the mean of realizations curves with the reference PN curves. Therefore, the PNSM could also provide a representative large PN with respect to two-phase flow properties when it is fed with signature parts.

It is important to emphasize that we have deliberately use a constructed large heterogeneous sample that is still small enough so that we are able to extract its PN in order to have a valid reference sample. However in practice, a reference PN of an entire core cannot be available in a real practical problem due to limitations as discussed in Section \ref{intro} and experimental measurements of rock flow properties can be used as a reference. In addition, this defined problem helped us to have properties of all different pieces of the sample in a way that we could readily choose three distinct ones as signature parts and use them as inputs of the developed PNSM. However, this step may not be as easy if we cannot capture properties of pieces of the entire core, and other methods such as fast pore size distribution evaluation on 2D industrial-scale images can be used to find those signature parts, and then, take micro-CT scans of them to get to signature PNs.

\section{Summary and Conclusions}
\label{conclusion}
We developed a pore-network stitching method (PNSM) to extend existing PN extraction and modeling capabilities to incorporate core-scale heterogeneity. The novel PNSM workflow takes a few signature parts representative of different heterogeneous parts of the core as inputs and use their extracted PNs and statistics of their geometrical and topological information to generate and stitch new pore elements in the empty space of the domain. The PNSM gives multiple realizations of a large PN with non-stationary statistical properties that vary in the volume generated between the signature PNs. We carried out validation studies on all steps of the workflow including layered stitching (in longitudinal and lateral directions) and volumetric stitching for different types of rock samples to demonstrate that the developed method is robust and independent of the morphology of the pore space. Then, we applied the method to two large domain heterogeneous problems and computed absolute permeability and drainage relative permeability curves of CO\textsubscript{2}-brine flow; this demonstrates that the method can construct a large representative PN for a heterogeneous porous medium and predict single-phase and two-phase flow properties successfully. Additional future work is necessary to apply the method on comprehensive multi-scale CT scans from a heterogeneous core and compare the predicted flow properties against core-scale experimental measurements.


\begin{acknowledgements}
The work was supported by the Center for Geologic Storage of CO\textsubscript{2}, an Energy Frontier Research Center funded by the U.S. Department of Energy (DOE), Office of Science, Basic Energy Sciences (BES), under Award \# DE-SC0C12504. 
\end{acknowledgements}

\bibliographystyle{apalike}
\bibliography{manuscript} 

\end{document}


\maketitle
\vspace{.50 in}

\section*{Validation on more samples}
In this document, we present two-phase flow pore-network simulation results on introduced samples in the Validation section of the paper. The samples are Berea sandstone (BR), Bentheimer sandstone (BN), Mt. Simon sandstone (ML), C1 carbonate (C1), C2 carbonate (C2). Basic information of their 3D images and properties are reported in the paper.

For each sample, we present the 3D reconstructed geometry of the rock and the segmented image of one slice of the stack. Then, we report drainage relative permeability curves of CO\textsubscript{2} and brine together for longitudinal and lateral layered stitching studies, respectively. Finally, validation of volumetric stitching are presented where relative permeability curves of 10 realizations and their mean are reported for CO\textsubscript{2} and brine phases, respectively. In all cases, the reference pore-network is obtained as discussed in the paper.

\begin{figure}[!htbp]
\centering
\includegraphics[width=0.50\textwidth]{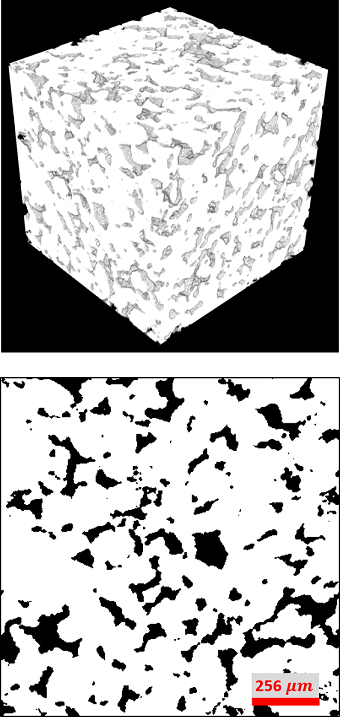}
\caption{The sample \textbf{BR}: (top) the 3D reconstructed geometry of the rock and (bottom) the segmented image of one slice of the stack.}
\label{fig_sample-br}
\end{figure}
\begin{figure}[!htbp]
\centering
\includegraphics[width=0.80\textwidth]{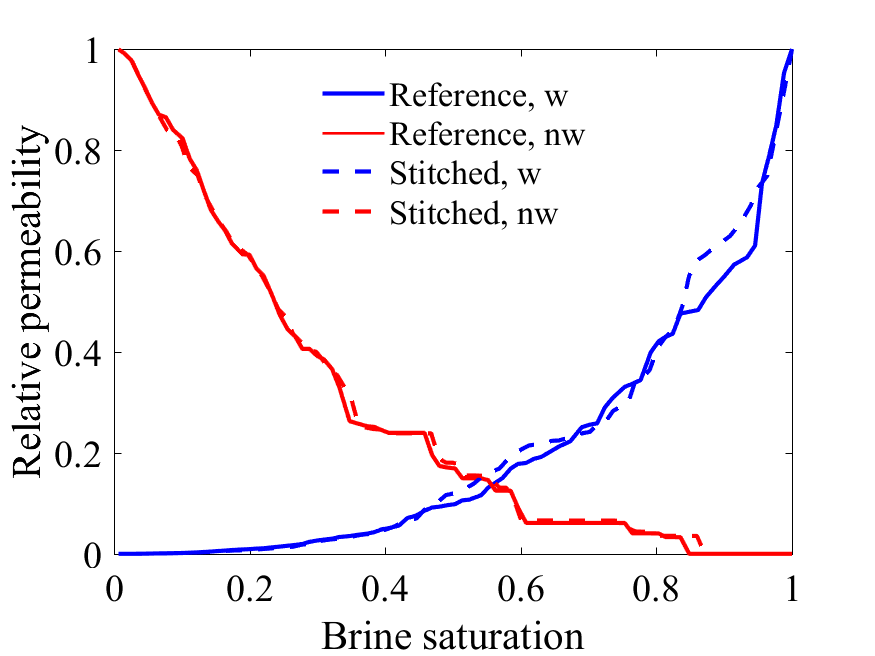}
\caption{Comparison of relative permeability curves of the stitched pore-network from sample \textbf{BR} via longitudinal layered stitching with its reference pore-network.}
\label{fig_val-laylong_br-kr}
\end{figure}
\begin{figure}[!htbp]
\centering
\includegraphics[width=0.80\textwidth]{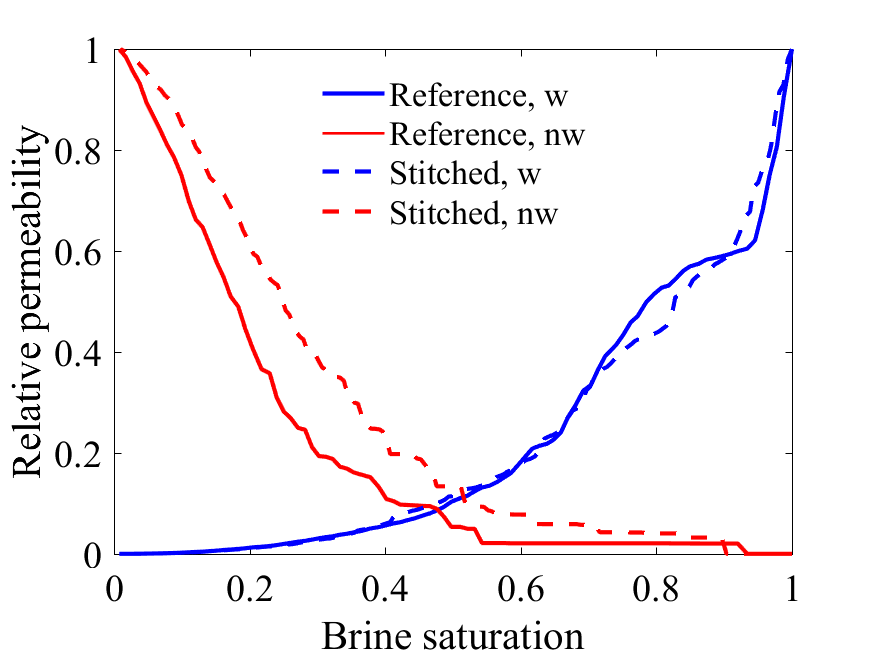}
\caption{Comparison of relative permeability curves of the stitched pore-network from sample \textbf{BR} via lateral layered stitching with its reference pore-network.}
\label{fig_val-laylat_br-kr}
\end{figure}
\begin{figure}[!htbp]
\centering
\includegraphics[width=0.80\textwidth]{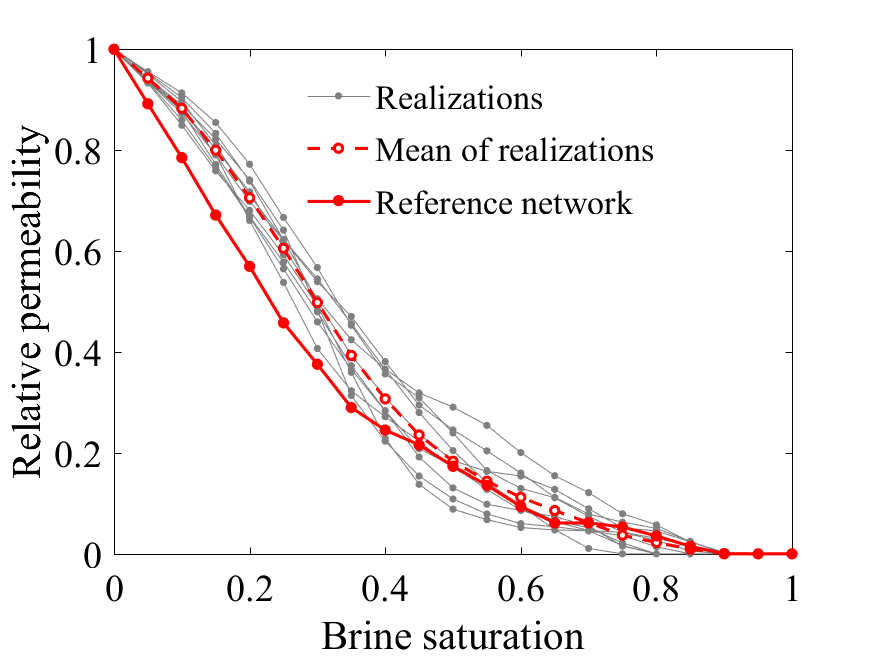}
\caption{Comparison of CO\textsubscript{2} relative permeability of 10 realizations of stitched pore-networks from sample \textbf{BR} via volumetric stitching with their reference pore-network.}
\label{fig_val-vol_br-krn}
\end{figure}
\begin{figure}[!htbp]
\centering
\includegraphics[width=0.80\textwidth]{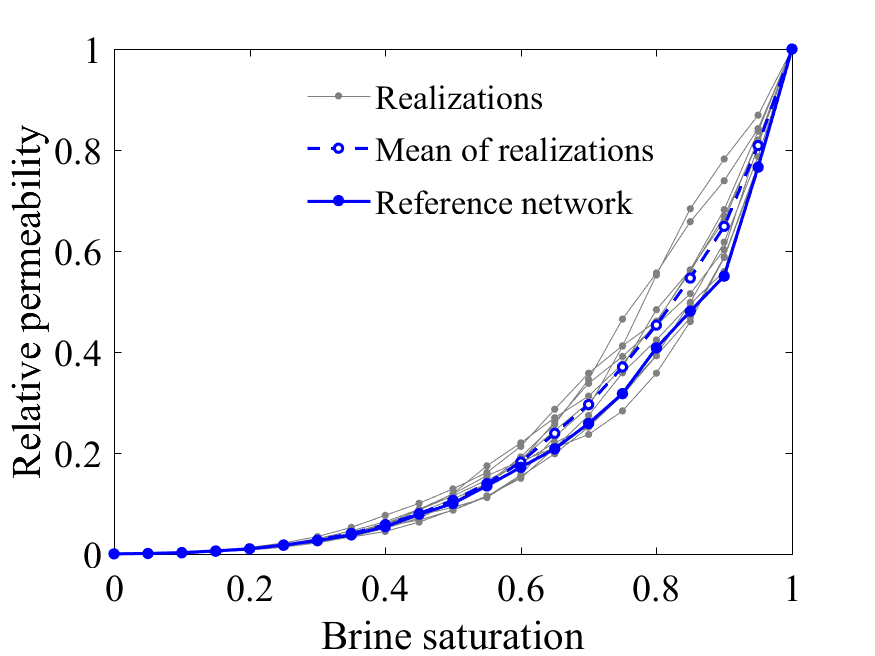}
\caption{Comparison of brine relative permeability of 10 realizations of stitched pore-networks from sample \textbf{BR} via volumetric stitching with their reference pore-network.}
\label{fig_val-vol_br-krw}
\end{figure}

\begin{figure}[!htbp]
\centering
\includegraphics[width=0.50\textwidth]{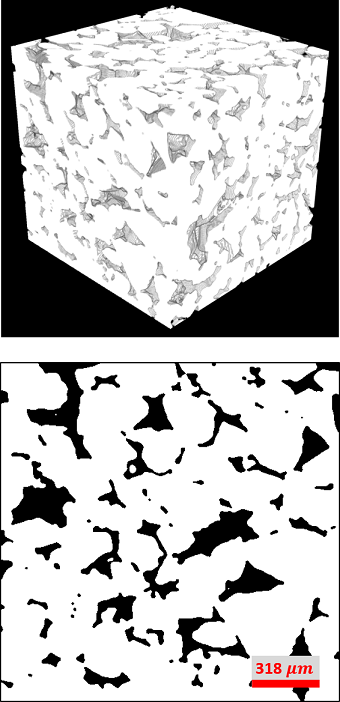}
\caption{The sample \textbf{BN}: (top) the 3D reconstructed geometry of the rock and (bottom) the segmented image of one slice of the stack.}
\label{fig_sample-bn}
\end{figure}
\begin{figure}[!htbp]
\centering
\includegraphics[width=0.80\textwidth]{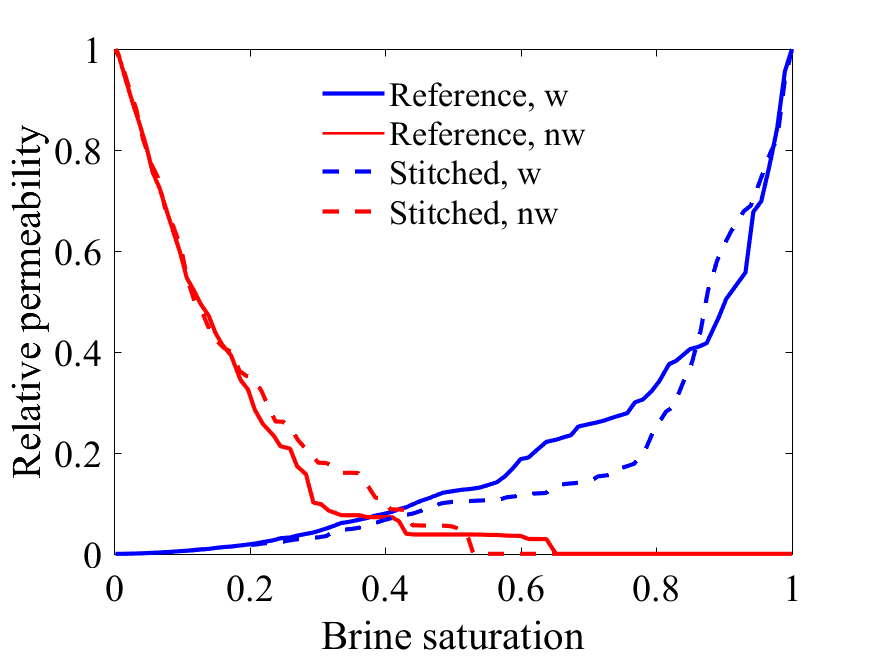}
\caption{Comparison of relative permeability curves of the stitched pore-network from sample \textbf{BN} via longitudinal layered stitching with its reference pore-network.}
\label{fig_val-laylong_bn-kr}
\end{figure}
\begin{figure}[!htbp]
\centering
\includegraphics[width=0.80\textwidth]{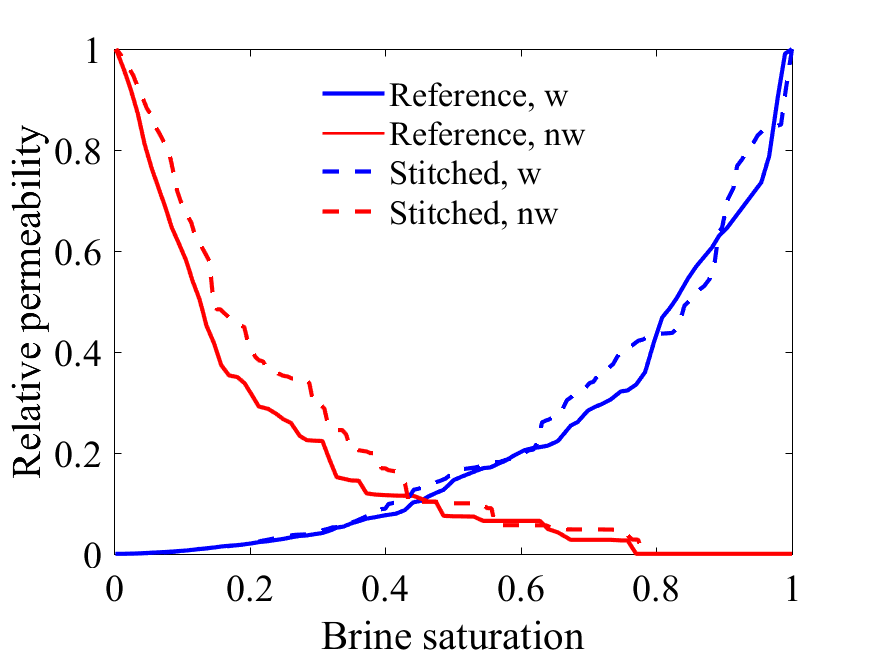}
\caption{Comparison of relative permeability curves of the stitched pore-network from sample \textbf{BN} via lateral layered stitching with its reference pore-network.}
\label{fig_val-laylat_bn-kr}
\end{figure}
\begin{figure}[!htbp]
\centering
\includegraphics[width=0.80\textwidth]{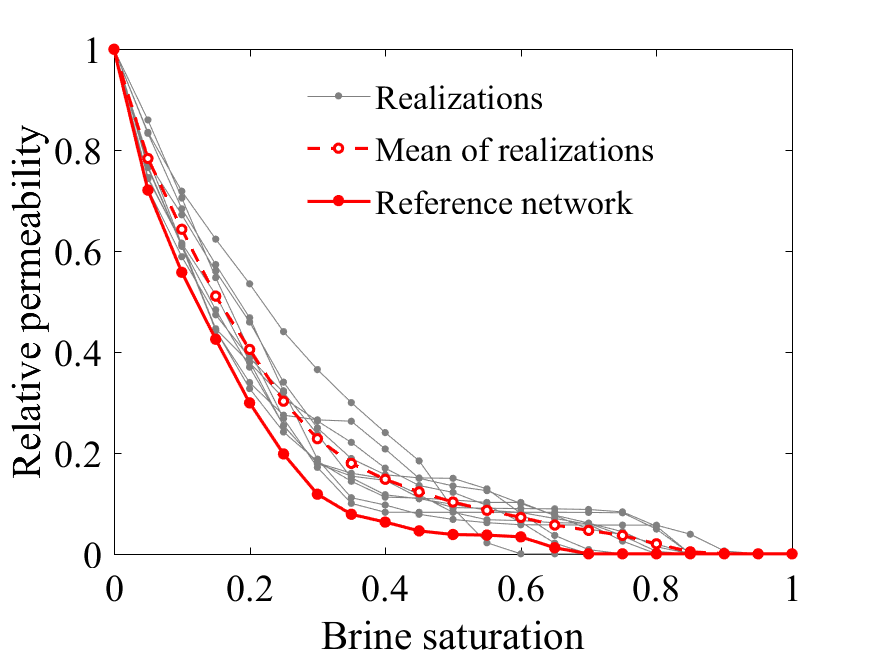}
\caption{Comparison of CO\textsubscript{2} relative permeability of 10 realizations of stitched pore-networks from sample \textbf{BN} via volumetric stitching with their reference pore-network.}
\label{fig_val-vol_bn-krn}
\end{figure}
\begin{figure}[!htbp]
\centering
\includegraphics[width=0.80\textwidth]{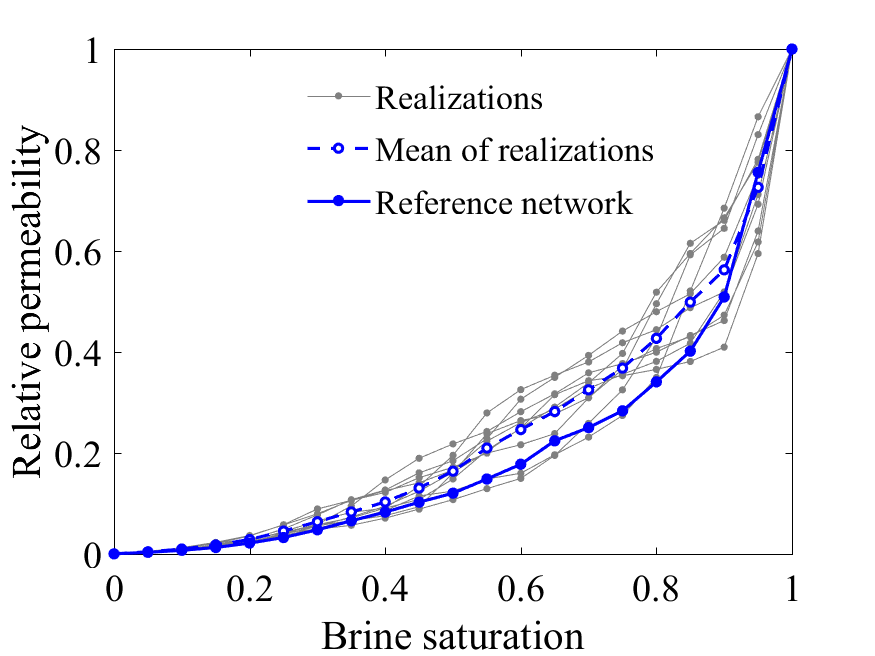}
\caption{Comparison of brine relative permeability of 10 realizations of stitched pore-networks from sample \textbf{BN} via volumetric stitching with their reference pore-network.}
\label{fig_val-vol_bn-krw}
\end{figure}

\begin{figure}[!htbp]
\centering
\includegraphics[width=0.50\textwidth]{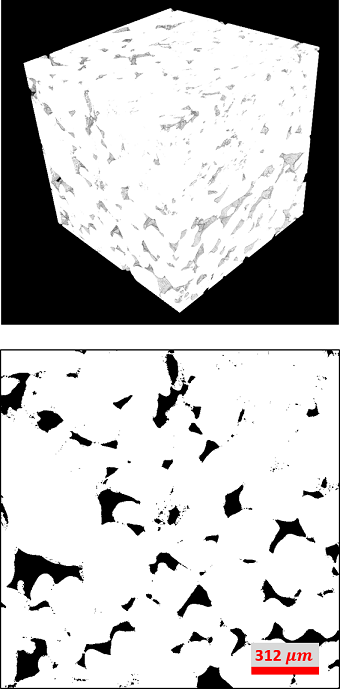}
\caption{The sample \textbf{ML}: (top) the 3D reconstructed geometry of the rock and (bottom) the segmented image of one slice of the stack.}
\label{fig_sample-ml}
\end{figure}
\begin{figure}[!htbp]
\centering
\includegraphics[width=0.80\textwidth]{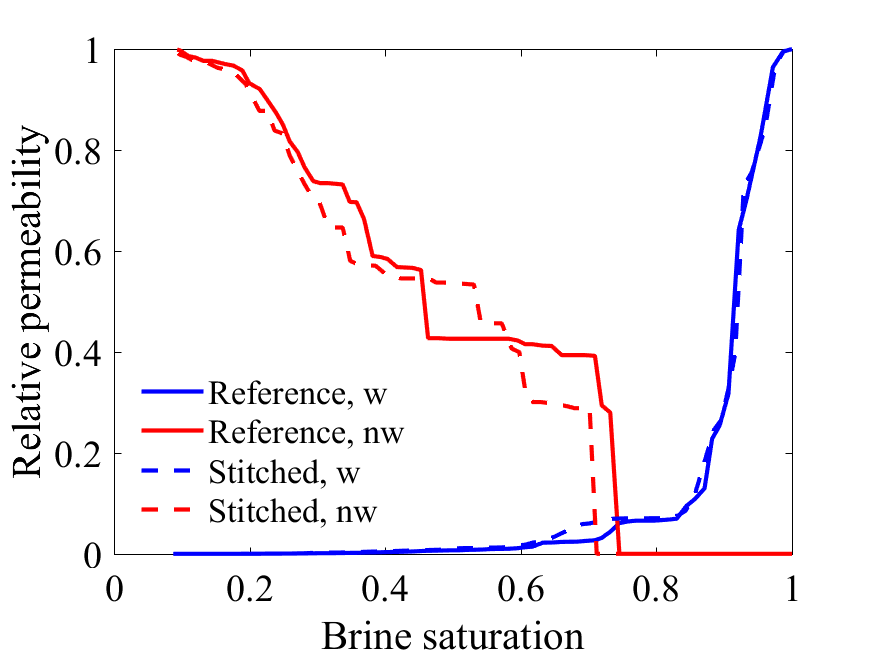}
\caption{Comparison of relative permeability curves of the stitched pore-network from sample \textbf{ML} via longitudinal layered stitching with its reference pore-network.}
\label{fig_val-laylong_ml-kr}
\end{figure}
\begin{figure}[!htbp]
\centering
\includegraphics[width=0.80\textwidth]{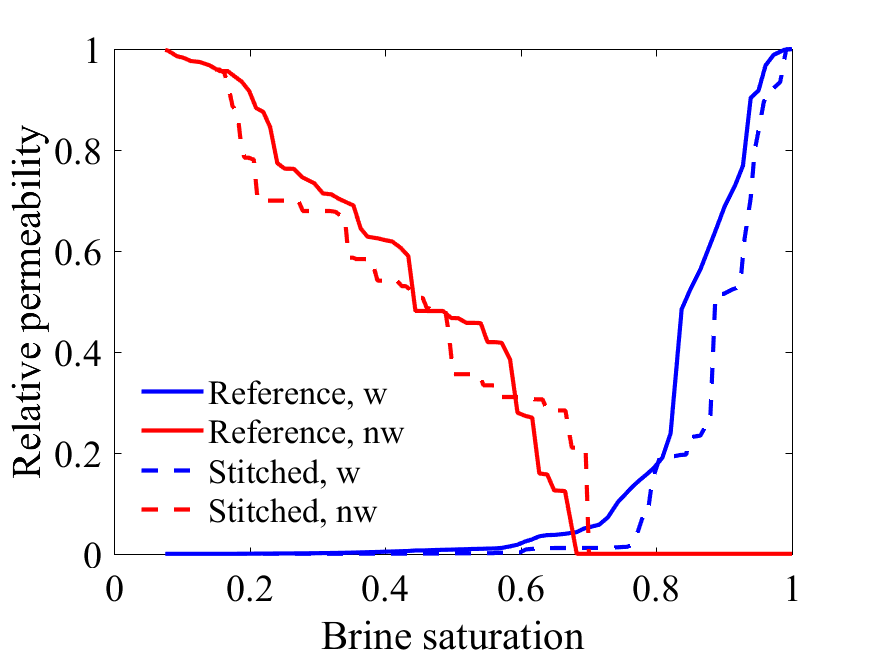}
\caption{Comparison of relative permeability curves of the stitched pore-network from sample \textbf{ML} via lateral layered stitching with its reference pore-network.}
\label{fig_val-laylat_ml-kr}
\end{figure}
\begin{figure}[!htbp]
\centering
\includegraphics[width=0.80\textwidth]{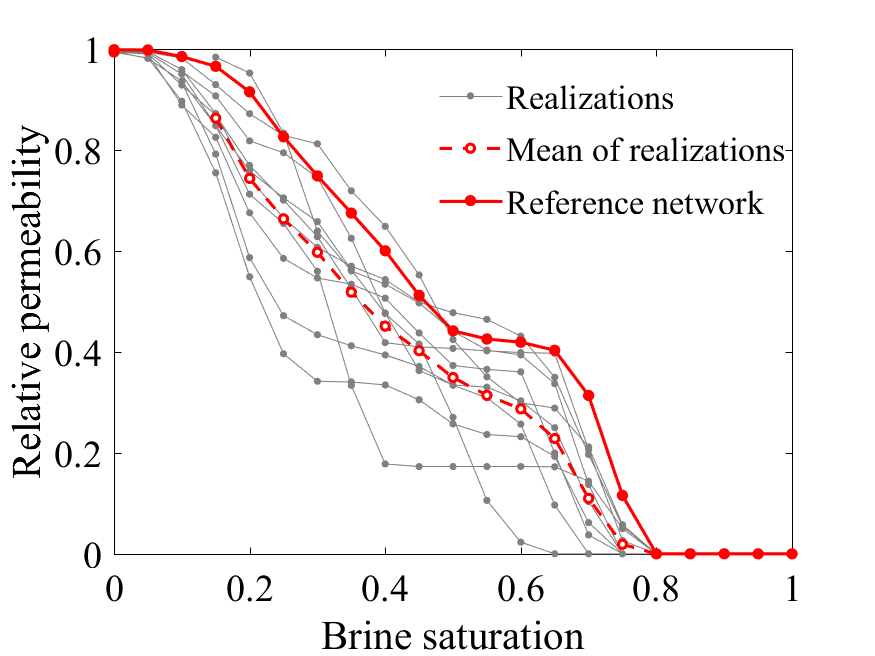}
\caption{Comparison of CO\textsubscript{2} relative permeability of 10 realizations of stitched pore-networks from sample \textbf{ML} via volumetric stitching with their reference pore-network.}
\label{fig_val-vol_ml-krn}
\end{figure}
\begin{figure}[!htbp]
\centering
\includegraphics[width=0.80\textwidth]{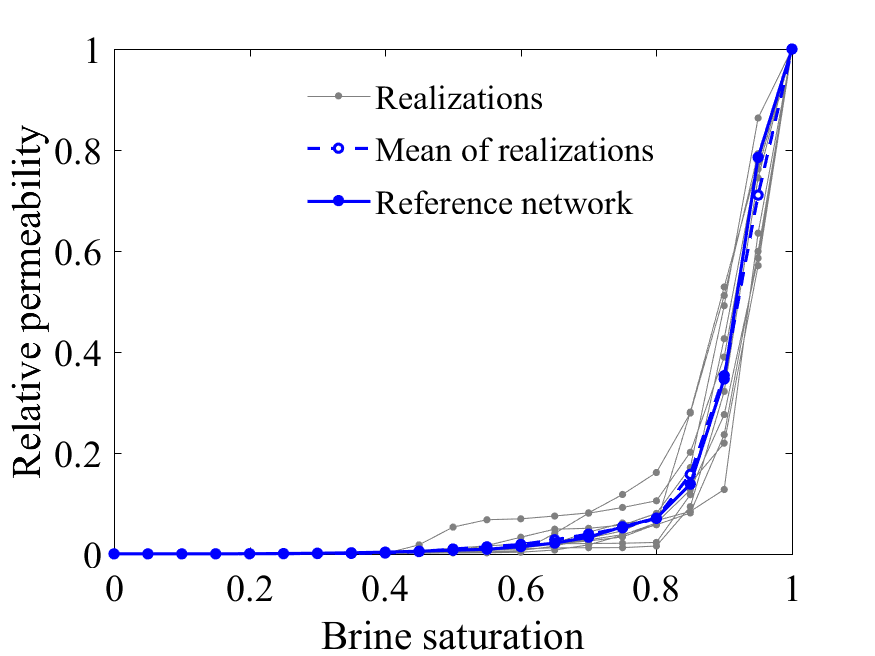}
\caption{Comparison of brine relative permeability of 10 realizations of stitched pore-networks from sample \textbf{ML} via volumetric stitching with their reference pore-network.}
\label{fig_val-vol_ml-krw}
\end{figure}

\begin{figure}[!htbp]
\centering
\includegraphics[width=0.50\textwidth]{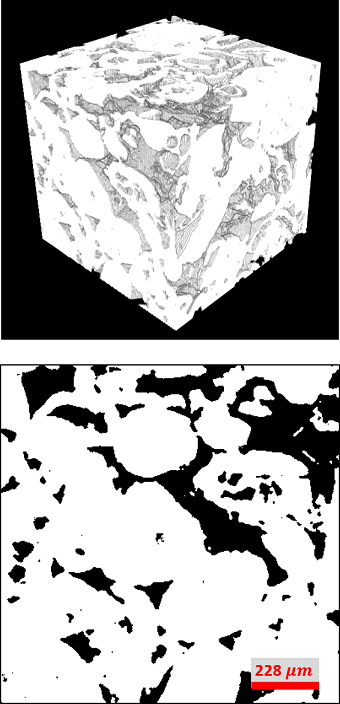}
\caption{The sample \textbf{C1}: (top) the 3D reconstructed geometry of the rock and (bottom) the segmented image of one slice of the stack.}
\label{fig_sample-c1}
\end{figure}
\begin{figure}[!htbp]
\centering
\includegraphics[width=0.80\textwidth]{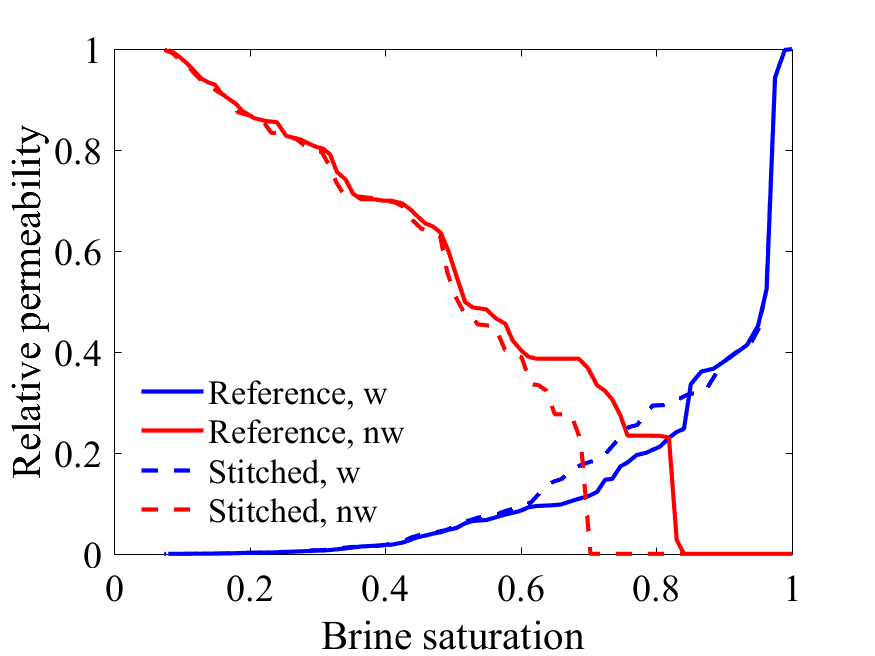}
\caption{Comparison of relative permeability curves of the stitched pore-network from sample \textbf{C1} via longitudinal layered stitching with its reference pore-network.}
\label{fig_val-laylong_c1-kr}
\end{figure}
\begin{figure}[!htbp]
\centering
\includegraphics[width=0.80\textwidth]{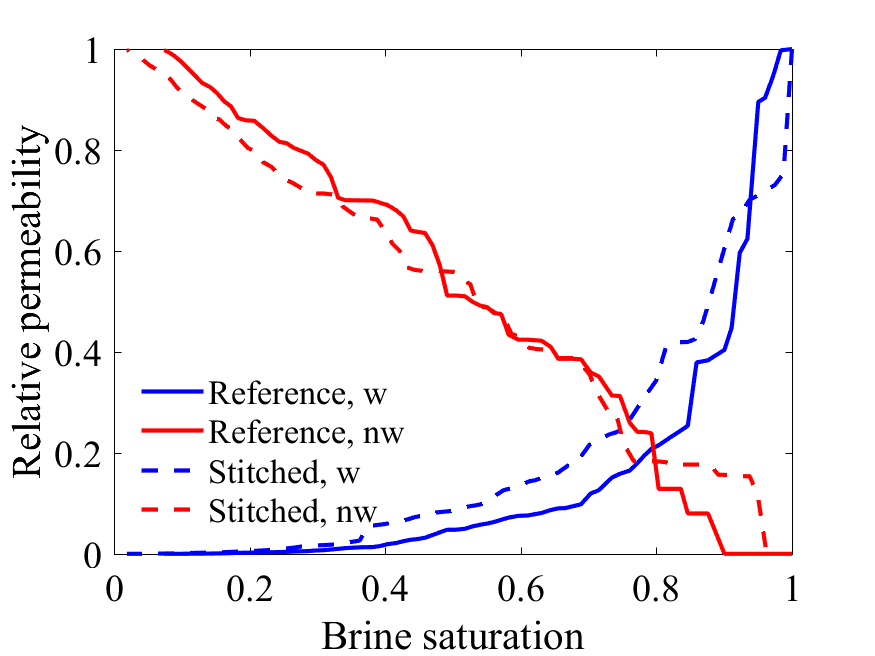}
\caption{Comparison of relative permeability curves of the stitched pore-network from sample \textbf{C1} via lateral layered stitching with its reference pore-network.}
\label{fig_val-laylat_c1-kr}
\end{figure}
\begin{figure}[!htbp]
\centering
\includegraphics[width=0.80\textwidth]{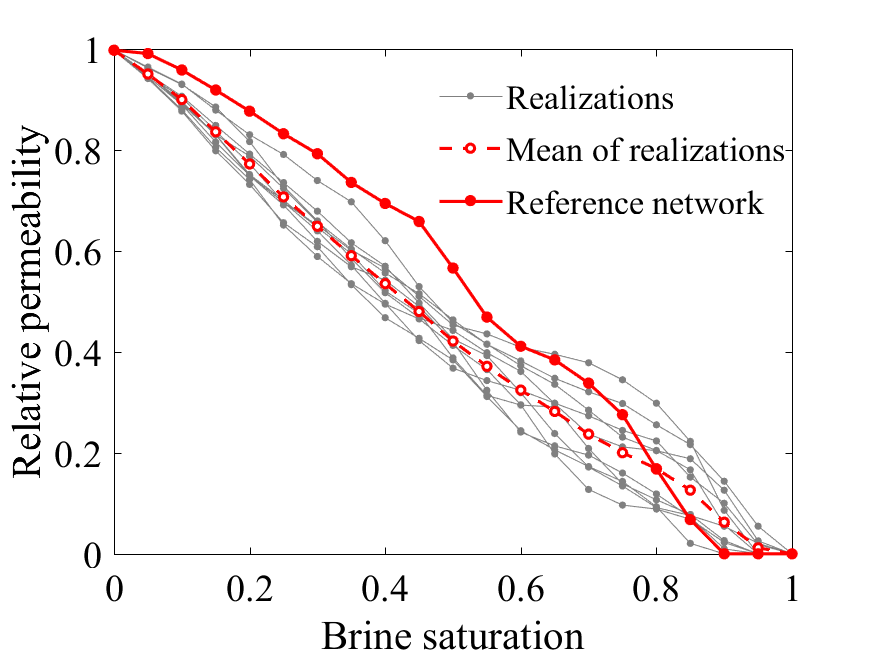}
\caption{Comparison of CO\textsubscript{2} relative permeability of 10 realizations of stitched pore-networks from sample \textbf{C1} via volumetric stitching with their reference pore-network.}
\label{fig_val-vol_c1-krn}
\end{figure}
\begin{figure}[!htbp]
\centering
\includegraphics[width=0.80\textwidth]{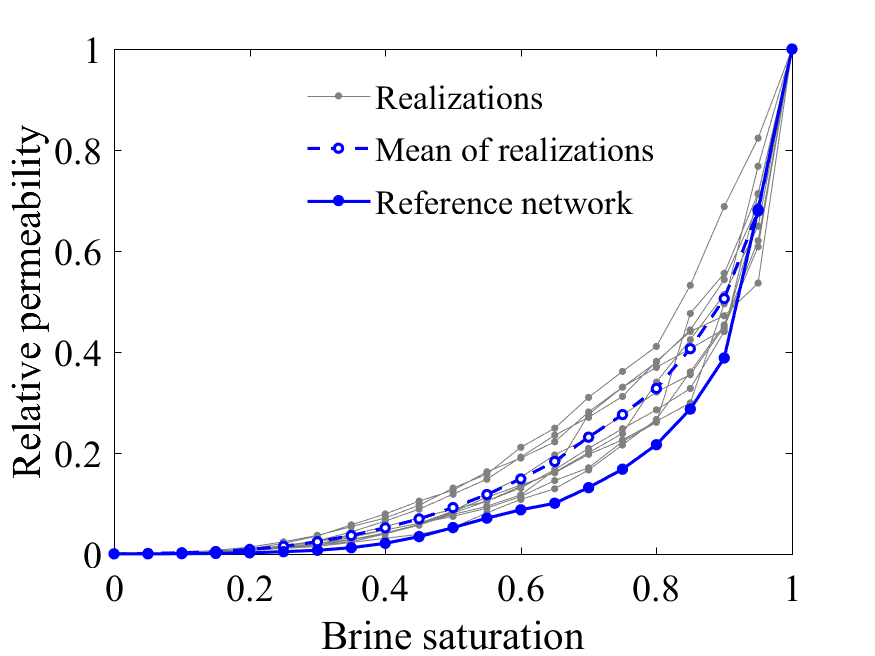}
\caption{Comparison of brine relative permeability of 10 realizations of stitched pore-networks from sample \textbf{C1} via volumetric stitching with their reference pore-network.}
\label{fig_val-vol_c1-krw}
\end{figure}

\begin{figure}[!htbp]
\centering
\includegraphics[width=0.50\textwidth]{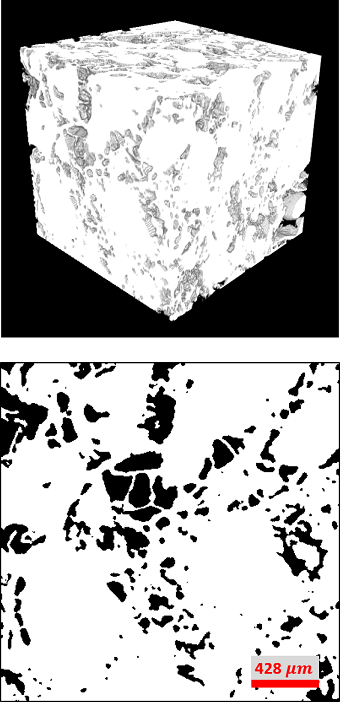}
\caption{The sample \textbf{C2}: (top) the 3D reconstructed geometry of the rock and (bottom) the segmented image of one slice of the stack.}
\label{fig_sample-c2}
\end{figure}
\begin{figure}[!htbp]
\centering
\includegraphics[width=0.80\textwidth]{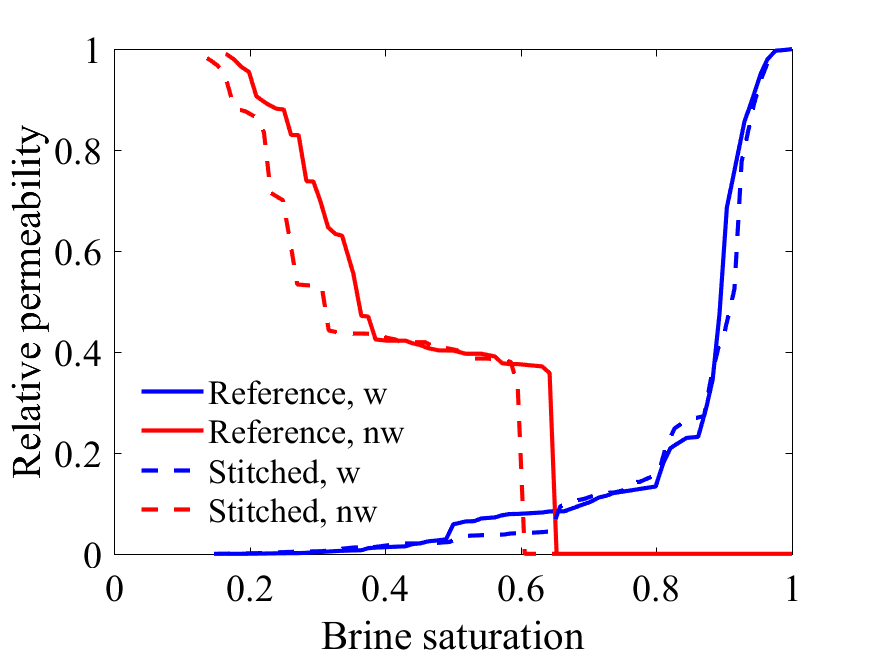}
\caption{Comparison of relative permeability curves of the stitched pore-network from sample \textbf{C2} via longitudinal layered stitching with its reference pore-network.}
\label{fig_val-laylong_c2-kr}
\end{figure}
\begin{figure}[!htbp]
\centering
\includegraphics[width=0.80\textwidth]{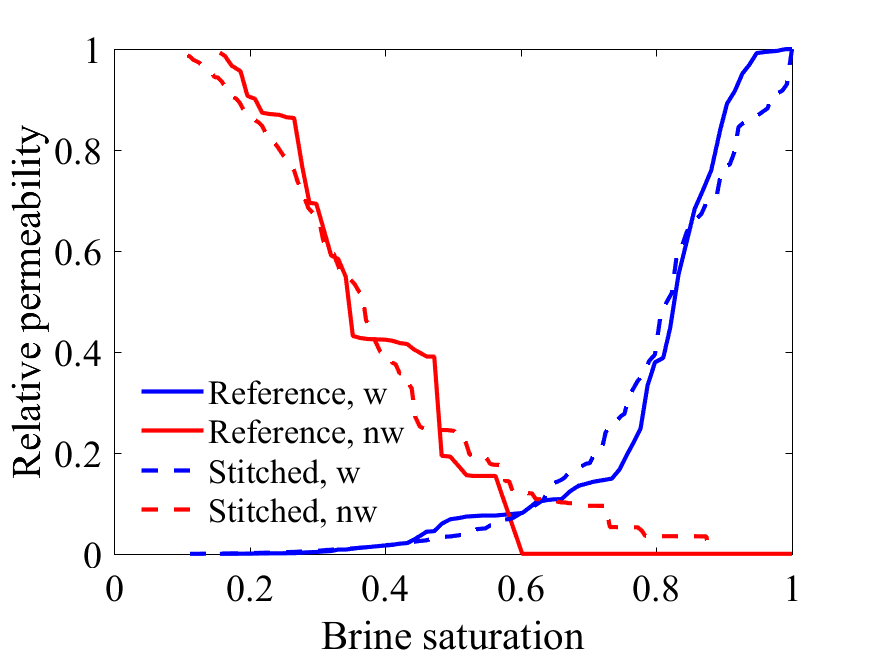}
\caption{Comparison of relative permeability curves of the stitched pore-network from sample \textbf{C2} via lateral layered stitching with its reference pore-network.}
\label{fig_val-laylat_c2-kr}
\end{figure}
\begin{figure}[!htbp]
\centering
\includegraphics[width=0.80\textwidth]{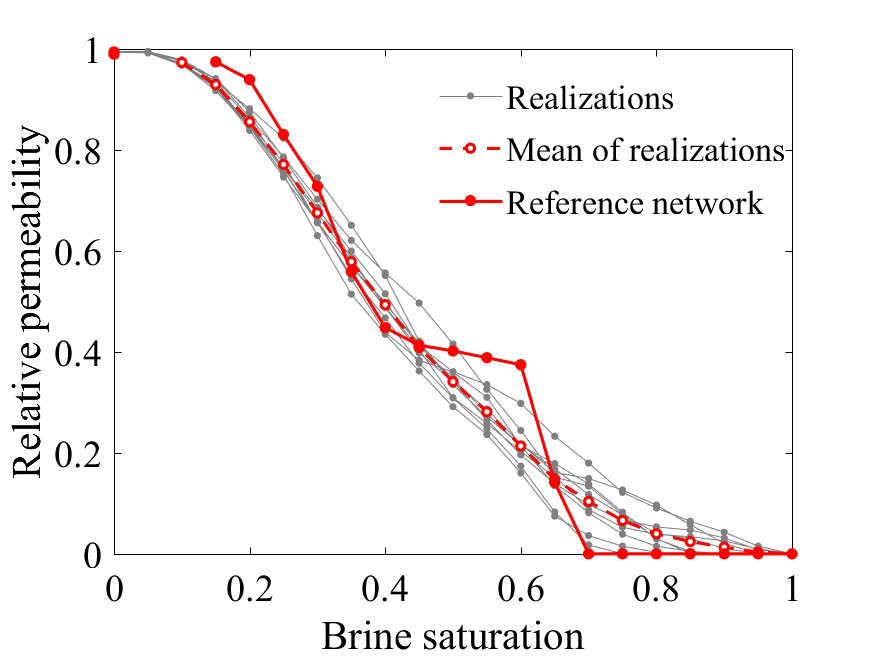}
\caption{Comparison of CO\textsubscript{2} relative permeability of 10 realizations of stitched pore-networks from sample \textbf{C2} via volumetric stitching with their reference pore-network.}
\label{fig_val-vol_c2-krn}
\end{figure}
\begin{figure}[!htbp]
\centering
\includegraphics[width=0.80\textwidth]{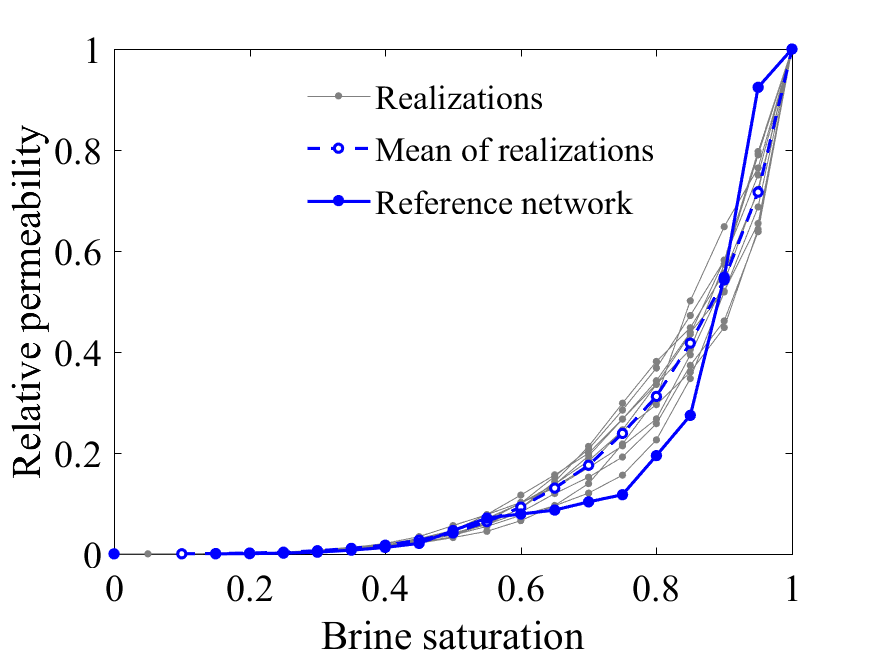}
\caption{Comparison of brine relative permeability of 10 realizations of stitched pore-networks from sample \textbf{C2} via volumetric stitching with their reference pore-network.}
\label{fig_val-vol_c2-krw}
\end{figure}